\documentclass[footinbib,aps,prb,reprint,twocolumn,superscriptaddress,groupedaddress,a4paper]{revtex4}

\usepackage{amsmath}
\usepackage{graphicx}
\usepackage{cleveref}
\usepackage{multirow}
\usepackage{amssymb}
\usepackage{mathtools}
\usepackage{color}
\usepackage{amsfonts}
\usepackage{epsfig}

\begin{document}


\widetext
\title{Impact of stoichiometry and strain on Ge$_{1-x}$Sn$_{x}$ alloys\\ from first principles calculations}

\author{Conor O'Donnell}
\affiliation{Tyndall National Institute, University College Cork, Lee Maltings, Dyke Parade, Cork T12 R5CP, Ireland}

\author{Alfonso Sanchez-Soares}
\affiliation{EOLAS Designs, Grenagh, Co.~Cork T23 AK70, Ireland}

\author{Christopher A.~Broderick}
\affiliation{Tyndall National Institute, University College Cork, Lee Maltings, Dyke Parade, Cork T12 R5CP, Ireland}
\affiliation{Department of Physics, University College Cork, Cork T12 YN60, Ireland}

\author{James C.~Greer}
\email{jim.greer@nottingham.edu.cn}
\affiliation{Nottingham Ningbo New Materials Institute and Department of Electrical and Electronic Engineering, University of Nottingham Ningbo China, 199 Taikang East Road, Ningbo 315100, China}

\date{\today}


\begin{abstract}

We calculate the electronic structure of germanium-tin (Ge$_{1-x}$Sn$_{x}$) binary alloys for $0 \leq x \leq 1$ using density functional theory (DFT). Relaxed alloys with semiconducting or semimetallic behaviour as a function of Sn composition $x$ are identified, and the impact of epitaxial strain is included by constraining supercell lattice constants perpendicular to the [001] growth direction to the lattice constants of Ge, zinc telluride (ZnTe), or cadmium telluride (CdTe) substrates. It is found that application of 1\% tensile strain reduces the Sn composition required to bring the (positive) direct band gap to zero by approximately 5\% compared to a relaxed Ge$_{1-x}$Sn$_{x}$ alloy having the same gap at $\Gamma$. On the other hand, compressive strain has comparatively less impact on the alloy band gap at $\Gamma$. Using DFT calculated alloy lattice and elastic constants, the critical thickness for Ge$_{1-x}$Sn$_{x}$ thin films as a function of $x$ and substrate lattice constant is estimated, and validated against supercell DFT calculations. The analysis correctly predicts the Sn composition range at which it becomes energetically favourable for Ge$_{1-x}$Sn$_{x}$/Ge to become amorphous. The influence of stoichiometry and strain is examined in relation to reducing the magnitude of the inverted (``negative'') $\Gamma_{7}^{-}$-$\Gamma_{8}^{+}$ band gap, which is characteristic of semimetallic alloy electronic structure. Based on our findings, strategies for engineering the semimetal-to-semiconductor transition via strain and quantum confinement in Ge$_{1-x}$Sn$_{x}$ nanostructures are proposed.

\end{abstract}

\maketitle


\section{Introduction}
\label{sec:introduction}

Germanium-tin (Ge$_{1-x}$Sn$_{x}$) alloys have attracted attention in recent years due to the indirect-to-direct band gap transition occurring as incorporation of Sn into Ge is increased to $x \approx 0.06 - 0.1$. The emergence of a direct band gap, combined with the reduction of the band gap energy with increasing Sn composition $x$, has motivated recent investigations of bulk-like Ge$_{1-x}$Sn$_{x}$, \cite{DarmodyEttisserryGoldsmanEtAl2015,SauCohen2007,DuttLinSukhdeoEtAl2013,Low2012} and Ge$_{1-x}$Sn$_{x}$ nanostructures. \cite{Doherty_CM_2020} Applications of Ge$_{1-x}$Sn$_{x}$ alloys in photonic devices, including semiconductor lasers and light-emitting diodes, \cite{Homewood2015,Wirths2015,Kasper_2015} and in electronic devices, including tunnelling field-effect transistors, \cite{Yang_IEEEIEDM_2012,Liu2016SimulationOG,Sant_IEEEJEDS_2015,Dunne2020} have been proposed and investigated. A motivation driving much of this interest is the potential for greater compatibility for on-chip photonic device integration with silicon-based microelectronics offered by use of a group-IV alloy.

Theoretical investigations have shown that the Ge$_{1-x}$Sn$_{x}$ band gap closes for Sn compositions $0.20 \lesssim x \lesssim 0.40$, with alloys at higher $x$ being inverted-gap semimetals. \cite{MoontragoonIkonicHarrison2007,JenkinsDow1987,Polak2017} This provides motivation to study Ge$_{1-x}$Sn$_{x}$ alloys at higher Sn compositions for potential applications in novel electronic devices. Here, we consider the potential to engineer the semimetal-to-semiconductor transition in Ge$_{1-x}$Sn$_{x}$ to facilitate the formation of monomaterial Schottky-like junctions, for applications in confinement-modulated-gap electronics. \cite{Ansari2012,nwt2017} The objective is to exploit quantum confinement effects in semimetallic thin films, nanowires or two-dimensional layers to create a thick semimetallic region abutting a thin semiconducting region. \cite{DLS12,Ansari2012,Sanchez-Soares2016,GAL17,GAK18} In the latter region, the nanostructure thickness is chosen below the length required to induce the opening of a band gap via quantum confinement, thereby allowing formation of a Schottky-like barrier within a single material. \cite{Ansari2012,GBA18} Notably, for such a junction no impurity doping is required, thereby eliminating fabrication challenges such as dopant segregation and dopant fluctuations on the few nanometre scale, nor is their a requirement to create a heterojunction. Electronic devices can be engineered with critical dimensions just above and below the critical length for quantum confinement to create rectifying diodes from the thick/thin junction's rectifying characteristics, or transistors by gating back-to-back thick/thin junctions.\cite{Ansari2012}

Diamond-structured $\alpha$-Sn is a group-IV semimetal, with its crystal structure and electronic properties making it promising for electronics applications. However, the use of semimetallic $\alpha$-Sn for device engineering presents two severe challenges from a practical perspective. The first of these is due to the structural instability of the $\alpha$ phase for temperatures $> 13.2$ $^\circ$C at ambient pressure, where it undergoes a phase transition to the tetragonal, metallic $\beta$ phase. \cite{Cohen1935, Houben_PRB_2019} Growth of lattice-matched $\alpha$-Sn thin films having thicknesses up to 500 nm has been reported, \cite{Farrow1981,Bowman1990} with the temperature at which the $\alpha$-Sn to $\beta$-Sn phase transition occurs increasing to approximately 70$^{\circ}$C. \cite{Farrow1981} However, this temperature is not sufficiently high for transistor operating conditions, which can exceed 100$^{\circ}$C. Additionally, the higher phase transition temperature remains far less than temperatures required during conventional integrated circuit fabrication. A second challenge is presented by the ``inverted'' band structure of $\alpha$-Sn, \cite{Groves1963} which admits a large separation of $> 0.5$ eV between the $\Gamma_{7}^{-}$ and $\Gamma_{8}^{+}$ states. This large ``negative'' $\Gamma_{7}^{-}$-$\Gamma_{8}^{+}$ band gap must be overcome by the confinement energy in a thin film or nanowire to open a band gap, \cite{Kuefner2013} which is required for creating a thick/thin junction for use as a Schottky-like barrier. Furthermore, the growth of thin films is in general constrained by critical thickness limitations in the presence of substrate lattice mismatch, presenting further challenges for the growth of structurally stable $\alpha$-Sn films. Here, we explore Ge$_{1-x}$Sn$_{x}$ alloys as a candidate material system to overcome these limitations in relation to confinement-modulated-gap electronic devices. \cite{Ansari2012}

 The evolution of the Ge$_{1-x}$Sn$_{x}$ band gap with increasing Sn composition -- from an indirect-gap semiconductor in Ge to an inverted-direct-gap semimetal in $\alpha$-Sn -- suggests a mechanism for engineering the length scale at which the onset of a confinement-induced semimetal-to-semiconductor transition occurs. In the absence of strain, incorporation of Ge in $\alpha$-Sn reduces the magnitude of the inverted $\Gamma_{7}^{-}$-$\Gamma_{8}^{+}$ energy gap. Reducing the magnitude of this negative band gap therefore relaxes constraints on nanostructure sizes required to open a band gap via quantum confinement. Predictions as to the precise nature of the inverted energy gap vary, being direct or indirect, or whether the zone-centre band ordering is topologically trivial or non-trivial. \cite{Lan2017,Polak2017} Here, we quantify the impact of alloy composition and strain in reducing the magnitude of the negative band gap in semimetallic Ge$_{1-x}$Sn$_{x}$. Ge incorporation in $\alpha$-Sn also enhances thermal lattice stability compared to $\alpha$-Sn, with the potential for materials processing at higher temperatures. \cite{Farrow1981,Ewald1954}

Due to the low solid solubility of Sn in Ge,\cite{OlesinskiAbbaschian1984} growth of Ge$_{1-x}$Sn$_{x}$ alloys with Sn compositions in the semimetallic range $x \gtrsim 0.20$ is challenging. Recent investigations have demonstrated that larger Sn compositions can be achieved in free-standing nanostructures due to enhanced strain relaxation relative to bulk-like epitaxial layers. \cite{Piao1990,HeAtwater1996,Biswas2016,Suzuki2016,Doherty_CM_2020} Sn incorporation has been demonstrated to correlate with thickness in thin film growth, with $x \approx 0.46$ achieved via growth of films having a thickness of $\approx 3$ nm. \cite{Suzuki2016} We have also recently quantified theoretically how miscibility can be increased via growth on lattice-matched substrates. \cite{Miscibility} Experimentally, growth at $x \approx 0.50$ has been achieved on gallium antimonide (GaSb) substrates. \cite{Piao1990}

Application of tensile strain has been predicted to induce an indirect- to direct-gap transition in Ge. \cite{Gupta2013,Virgilio2013,ElKurdi2010a} While application of high tensile strain can induce a semiconducting-to-semimetallic transition in Ge, the required strains -- corresponding to $\approx 4.5$\% lattice-mismatch in a pseudomorphically strained, [001]-oriented epitaxial layer \cite{ElKurdi2010a} -- are sufficiently large that critical thickness limitations make this level of strain impractical. However, the results of Ref.~\onlinecite{ElKurdi2010a} suggest that tensile strain can be exploited to reduce the Sn composition at which Ge$_{1-x}$Sn$_{x}$ becomes semimetallic. To provide a systematic analysis of the incorporation of higher Sn compositions in Ge$_{1-x}$Sn$_{x}$, and to determine the resulting impact on the electronic structure, we undertake first principles electronic structure calculations for pseudomorphically strained Ge$_{1-x}$Sn$_{x}$ as a function of both Sn composition and strain. The latter is investigated by pseudomorphically straining Ge$_{1-x}$Sn$_{x}$ alloy supercells with respect to Ge, ZnTe and CdTe substrates, with ZnTe and CdTe chosen as they respectively possess lattice constants close to that of the fictitious zinc blende IV-IV compound zb-GeSn, \cite{OHalloran2019,Tanner_submitted_2020} and of $\alpha$-Sn. To assess the feasibility of thin film growth at different strains and alloy stoichiometries, we also compute the critical thickness of pseudomorphically strained Ge$_{1-x}$Sn$_{x}$, thereby quantifying strain-related limitations to pseudomorphic growth. This allows a balance between achieving a desired electronic structure and the chosen stoichiometry and strain, while considering achievable epitaxial thin film thicknesses.

The remainder of this paper is organised as follows. In Sec.~\ref{sec:theory} we describe the theoretical methods employed in our analysis, focusing respectively in Secs.~\ref{sec:theory_dft},~\ref{sec:theory_sqs},~\ref{sec:theory_strain} and~\ref{sec:theory_critical} on DFT calculations, alloy special quasi-random structures (SQSs), the impact of pseudomorphic strain on the band structure, and critical thickness limitations. Our results are presented in Sec.~\ref{sec:results}, beginning in Sec.~\ref{sec:GeSn_unstrained} with our study of the evolution of the band gap of unstrained Ge$_{1-x}$Sn$_x$ alloys across the full alloy composition range. Sec.~\ref{sec:results_strained} is concerned with the impact of pseudomorphic strain on the band gap evolution in Ge$_{1-x}$Sn$_{x}$ grown on Ge, ZnTe and CdTe substrates. A comparison between the strained band gaps calculated by DFT and those calculated via deformation potential theory is presented in Sec.~\ref{sec:results_transition}, where we also describe the critical thickness of strained Ge$_{1-x}$Sn$_{x}$ grown on Ge, ZnTe or CdTe substrates, and quantify the ability to combine alloying and strain to engineer the composition at which the band gap energy becomes zero, demarcating the boundary between semiconducting and semimetallic strained alloys. Finally, in Sec.~\ref{sec:conclusions}, we summarise and conclude.


\section{Methodology}
\label{sec:theory}


\subsection{Electronic structure: density functional theory}
\label{sec:theory_dft}

Our analysis of the structural, elastic and electronic properties of Ge$_{1-x}$Sn$_{x}$ alloys is based on Kohn-Sham DFT. \cite{Dreizler1990d} Structural relaxations are carried out in the local density approximation (LDA). \cite{Perdew1981} Electronic band structures are calculated using the Tran-Blaha modified Becke-Johnson (TB-mBJ) meta-generalised gradient approximation (meta-GGA) exchange-correlation potential, \cite{Tran2009} in order to overcome the band gap underestimation typical of the LDA in the Kohn-Sham formalism. Due to large relativistic effects for Sn, our electronic structure calculations include spin-orbit coupling. However, spin-orbit coupling is neglected during structural relaxations, which reduces computational time with negligible differences in the final relaxed ionic positions. Our DFT calculations employ a basis of numerical atomic orbitals (NAOs), \cite{Soler2002,Ozaki2003,Ozaki2004} with a cut-off energy of 100 Ha for real-space integration. Reciprocal space integration is carried out using $\Gamma$-centred Monkhorst-Pack grids \cite{Monkhorst1976} with a density of $\geq 7$ \textbf{k}-points \AA$^{-1}$ along each reciprocal lattice vector. For Ge we employ a NAO basis consisting of four $s$, four $p$, three $d$ and two $f$ orbitals per atom ($s^{4}p^{4}d^{3}f^{2}$). We employ norm-conserving pseudopotentials in which the $(4s)^{2}$ and $(4p)^{2}$ states of Ge are treated explicitly as valence states. The semi-core $(3d)^{10}$ states of Ge are treated as core states, since unfreezing them has been demonstrated to have a minimal impact on the calculated electronic structure.\cite{Eckhardt_PRB_2014,OHalloran2019} For Sn we employ a $s^{4}p^{4}d^{3}f^{2}$ NAO basis, and norm-conserving pseudopotentials in which the $(4d)^{10}$, $(5s)^{2}$ and $(5p)^{2}$ orbitals are treated explicitly as valence states. The Becke-Roussel mixing parameter $c$ in the TB-mBJ exchange-correlation potential is treated as an adjustable parameter, and varied independently for Ge and Sn in order to accurately reproduce the overall band structure of Ge and $\alpha$-Sn. We use $c_{\scalebox{0.7}{\textrm{Ge}}} = 1.100$ and $c_{\scalebox{0.7}{\textrm{Sn}}} = 1.225$ for our TB-mBJ calculations. The accuracy of this approach to simulate the Ge$_{1-x}$Sn$_{x}$ electronic structure has been verified recently by several groups. \cite{Eckhardt_PRB_2014,Polak2017,OHalloran2019}

\begin{figure}
\centering
\includegraphics[width=0.80\columnwidth]{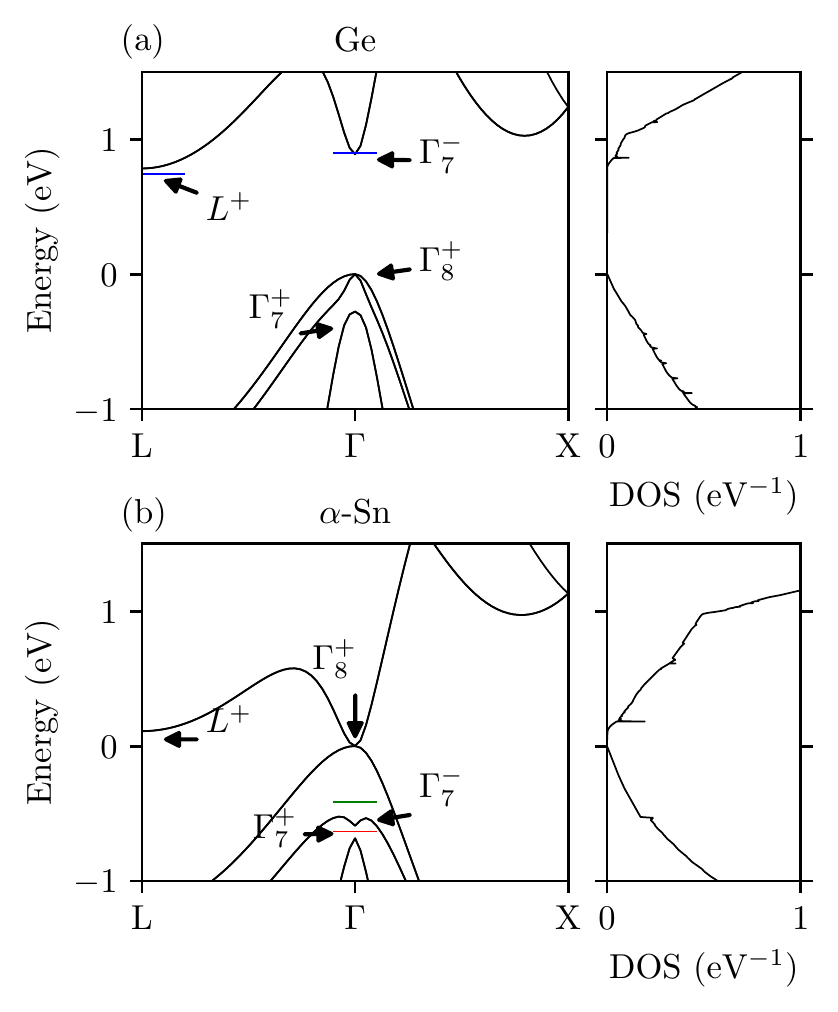}
\caption{DFT-calculated band structure (left-hand panel) and DOS (right-hand panel) of (a) Ge, and (b) $\alpha$-Sn. The zero of energy is chosen to lie at the Fermi energy. Note the inverted ordering of the $\Gamma_{8}^{+}$ and $\Gamma_{7}^{-}$ states for Sn relative to Ge. In (a) the solid blue lines denote the measured low-temperature indirect (fundamental) L$_{6}^{+}$-$\Gamma_{8}^{+}$ and direct $\Gamma_{7}^{-}$-$\Gamma_{8}^{+}$ band gaps of Ge. \cite{Zwerdling1959} In (b) the solid green and red lines denote the measured low temperature $\Gamma_{8}^{+}$-$\Gamma_{7}^{-}$ (inverted band gap) energy in $\alpha$-Sn. \cite{GROVES19702031,Booth1968_a}}
\label{fig:2a_Pure_exp}
\end{figure}

Figures~\ref{fig:2a_Pure_exp}(a) and~\ref{fig:2a_Pure_exp}(b) respectively show the band structure and density of states (DOS) of Ge and $\alpha$-Sn calculated using this approach. Comparison of the results of our DFT calculations to experimental data and to previous calculations is provided in Table~\ref{tab:dft_benchmark}, where we list the LDA-calculated lattice and elastic constants, as well as the TB-mBJ-calculated direct $\Gamma_{7}^{-}$-$\Gamma_{8}^{+}$ and indirect L$_{6}$-$\Gamma_{8}^{+}$ band gaps, the associated hydrostatic deformation potentials $a_{g} ( \Gamma_{7}^{-}\text{-}\Gamma_{8}^{+} )$ and $a_{g} ( \text{L}_{6}^{+}\text{-}\Gamma_{8}^{+} )$, the valence band (VB) spin-orbit splitting energy $\Delta_{0} = E( \Gamma_{8}^{+} ) - E( \Gamma_{7}^{+} )$, and the VB edge axial and shear deformation potentials $b$ and $d$.

For Ge$_{1-x}$Sn$_{x}$ alloy supercell TB-mBJ calculations, we interpolate the Becke-Roussel mixing parameter $c$ as
\begin{equation}
     c(x) = \frac{ ( 1 - x )\, \Omega_{\scalebox{0.7}{\textrm{Ge}}} \, c_{\scalebox{0.7}{\textrm{Ge}}} + x \, \Omega_{\scalebox{0.7}{\textrm{Sn}}}\, c_{\scalebox{0.7}{\textrm{Sn}}} }{ (1 - x) \, \Omega_{\scalebox{0.7}{\textrm{Ge}}}+  x \, \Omega_{\scalebox{0.7}{\textrm{Sn}}} },
     \label{eqn:mgga_c}
\end{equation}
\noindent
where $\Omega_{\scalebox{0.7}{\textrm{Sn}}}$ and $\Omega_{\scalebox{0.7}{\textrm{Ge}}}$ are the equilibrium volumes of $\alpha$-Sn and Ge primitive unit cells, respectively. Equation~\eqref{eqn:mgga_c} therefore gives the mixing parameter by weighting $c_{\scalebox{0.7}{\textrm{Ge}}}$ and $c_{\scalebox{0.7}{\textrm{Sn}}}$ based on the fractional volume each element occupies in an alloy supercell at Sn composition $x$, thereby weighting exchange-correlation effects based upon the difference in covalent radii of Ge and Sn.


\begin{table*}[t!]
	\caption{\label{tab:dft_benchmark} DFT-calculated lattice and elastic constants, and electronic properties of Ge and $\alpha$-Sn, compared to previous theoretical calculations and experimental measurements. The lattice and elastic constants $a$, $C_{11}$, $C_{12}$ and $C_{44}$ were computed via LDA-DFT. The direct and indirect band gaps, $E_{g}^{\Gamma} = E( \Gamma_{7}^{-} ) - E( \Gamma_{8}^{+} )$ and $E_{g}^{\protect\scalebox{0.7}{\text{L}}} = E( \text{L}_{6}^{+} ) - E( \Gamma_{8}^{+} )$, the VB spin-orbit splitting energy $\Delta_{0} = E( \Gamma_{8}^{+} ) - E( \Gamma_{7}^{+} )$, the direct- and indirect-gap hydrostatic deformation potentials $a_{g}( \Gamma_{7}^{-}$-$\Gamma_{8}^{+} )$ and $a_{g}( \text{L}_{6}^{+}$-$\Gamma_{8}^{+} )$, and the VB edge axial deformation potential $b( \Gamma_{8}^{+} )$ were computed via meta-GGA (TB-mBJ) DFT.}
	\begin{ruledtabular}
		\begin{tabular}{cc|ccc|ccc}
			                                               &          &              & Ge           &                           &             & $\alpha$-Sn  &              \\
			Parameter                                      & Unit     & This work    & Theory       & Experiment                & This work   & Theory       & Experiment   \\
			\hline
			$a$                                            & \AA      & 5.64         & 5.646$^{a}$  & 5.657$^{b}$               & 6.47        & 6.49$^{a}$   & 6.489$^{c}$  \\
			$C_{11}$                                       & GPa      & 122.96       & 122$^{d}$,   & 128.53$^{e}$,             & 68.27       & 68$^{d}$,    & 69$^{f}$             \\
						                                   &          &              & 142.5 $^{g}$ & 128.9$^{h}$               &             & 72.53$^{g}$  &   70$^{i}$           \\
			$C_{12}$                                       & GPa      & 49.75        & 47$^{d}$,    & 48.25$^{e}$,              & 36.59       & 34$^{d}$     &   29.3$^{f}$           \\
						                                   &          &              & 58.5$^{g}$   & 48.3$^{h}$                &             & 29.73$^{g}$  &    33$^{i}$          \\
			$C_{44}$                                       & GPa      & 60.83        & 86$^{d}$,    & 66.8$^{e}$,               & 28.98       & 53$^{d}$,    & 36.2$^{f}$             \\
						                                   &          &              & 58.7$^{g}$   & 67.1$^{h}$                &             & 29.9$^{g}$   &     32$^{i}$          \\
			\hline
			$E( \Gamma_{7}^{-}   ) - E( \Gamma_{8}^{+} )$  & eV       & 0.89         & 0.879$^{j}$, & 0.898$^{k}$               & -0.589      & -0.39$^{j}$, & -0.413$^{l}$,  \\
						                                   &          &              & 0.892$^{m}$  &                           &             & -0.408$^{m}$,& -0.634$^{n}$   \\
						                                   &          &              &              &                           &             & -0.64$^{o}$  &              \\
			$E( \text{L}_{6}^{+} ) - E( \Gamma_{8}^{+} )$  & eV       & 0.78         & 0.71$^{j}$   & 0.744$^{k}$               &             &              &               \\
						                                   &          &              & 0.744$^{m}$  &                           &             &              &              \\
			$E( \Gamma_{8}^{+}   ) - E( \Gamma_{7}^{+} )$  & eV       & 0.28         & 0.27$^{p}$   & 0.296$^{q}$               & 0.68        & 0.8$^{o}$    &  \\
						                                   &          &              & 0.3$^{o}$    &                           &             & 0.8$^{n}$    &              \\
			\hline
			$a_{g}( \Gamma_{7}^{-}$-$\Gamma_{8}^{+} )$       & eV       & -9.54        & -8.6$^{q}$   &                           & -6.68       & -9.1$^{q}$,  &  \\
						                                   &          &              &              &                           &             & -6.97$^{r}$  &              \\
			$a_{g}( \text{L}_{6}^{+}$-$\Gamma_{8}^{+} )$       & eV       & -3.36        & -2.78$^{s}$  &                           & -1.53       &              &                \\
			$b( \Gamma_{8}^{+} )$                          & eV       & -2.78        & -2.66$^{o}$  &                           & -2.39       & -2.31$^{o}$  &              \\
						                                   &          &              & -2.16$^{t}$  &                           &             &              &              \\
			$d( \Gamma_{8}^{+} )$			               & eV       & -5.95        & 6.06$^{t}$   &                           & -5.23       & -4.1$^{u}$   &              \\
		\end{tabular}
	\end{ruledtabular}
	\begin{flushleft}

	$^{a}$Ref.~\onlinecite{Gupta2013}\;
	$^{b}$Ref.~\onlinecite{Baker1975} \;
	$^{c}$Ref.~\onlinecite{Farrow1981, Thewlis1954}\;
	$^{d}$Ref.~\onlinecite{Polak2017}\;
	$^{e}$Ref.~\onlinecite{McSkimin1963}\;
	$^{f}$Ref.~\onlinecite{Price1971}\;
	$^{g}$Ref.~\onlinecite{Shen1994}\;
	$^{h}$Ref.~\onlinecite{Huntington1958}\;
	$^{i}$Ref.~\onlinecite{Zdetsis1977}\;
	$^{j}$Ref.~\onlinecite{MoontragoonIkonicHarrison2007}\;
	$^{k}$Ref.~\onlinecite{Zwerdling1959}\;
	$^{l}$Ref.~\onlinecite{Booth1968_a}\;
	$^{m}$Ref.~\onlinecite{WirthsBucaMantl2016}\;
	$^{n}$Ref.~\onlinecite{GROVES19702031}\;
	$^{o}$Ref.~\onlinecite{Qteish1992}\;
	$^{p}$Ref.~\onlinecite{Roedl}\;
	$^{q}$Ref.~\onlinecite{Authors2002}\;
	$^{r}$Ref.~\onlinecite{Wei1999}\;
	$^{s}$Ref.~\onlinecite{VandeWalle1989}\;
	$^{t}$Ref.~\onlinecite{Fischetti1996}\;
	$^{u}$Ref.~\onlinecite{Roman1972}\;

	\end{flushleft}
\end{table*}


\subsection{Strained special quasi-random supercells}
\label{sec:theory_sqs}

Our DFT calculations are computationally demanding and therefore impose practical limits to the atomistic supercell sizes that can be employed to treat a disordered alloy such as Ge$_{1-x}$Sn$_{x}$. We therefore use the special quasi-random structure (SQS) supercell approach. \cite{Zunger1990, Hass1990} A SQS is constructed by minimising the difference between supercell lattice correlation functions and those associated with a random binary alloy, with the goal of providing a best approximation to a random alloy using a finite supercell. Our calculations employ 64-atom ($2 \times 2 \times 2$ simple cubic) Ge$_{1-x}$Sn$_{x}$ SQSs, generated stochastically using a simulated annealing procedure as implemented in the Alloy Theoretic Automated Toolkit (\textsc{ATAT}). \cite{ATAT2002,Miscibility} To simulate [001]-oriented Ge$_{1-x}$Sn$_{x}$ strained epitaxial layers, we apply pseudomorphic strain by (i) restricting the supercell lattice constant in the plane perpendicular to the [001] growth direction to be equal to that of a chosen substrate material, and (ii) performing structural relaxation by allowing the lattice constant along the growth direction, and the internal atomic degrees of freedom (ionic positions), to relax freely to minimise the lattice free energy. \cite{Miscibility}

The use of SQS supercells results in (i) local relaxation of the crystal lattice, which breaks the underlying cubic symmetry of the diamond structure which for example would be retained in a virtual crystal approximation (VCA), and hence lifts band degeneracies present for high-symmetry points in the band structures of Ge and $\alpha$-Sn, \cite{VandeWalle1989,Krijn_SST_1991} and (ii) folding of the electronic band structures due to the reduced size of the supercell Brillouin zone compared to that associated with a primitive cubic cell of an ideal crystalline version of the alloy. These factors complicate interpretation of the alloy electronic structure. \cite{Polak2017,OHalloran2019} To overcome these limitations and track the evolution of the conduction band (CB) and VB edge states in SQS supercells, we proceed by calculating overlaps between alloy states and selected states of an unperturbed Ge$_{64}$ supercell. This approach, which is similar to the computation of the spectral function at a single \textbf{k}-point employed in popular zone unfolding schemes, enables the evolution of both the energy and character of the alloy band edge states to be identified and tracked (see, e.g., \cite{Usman_PRA_2018}).

\begin{figure}
\centering
\includegraphics[width=0.94\columnwidth]{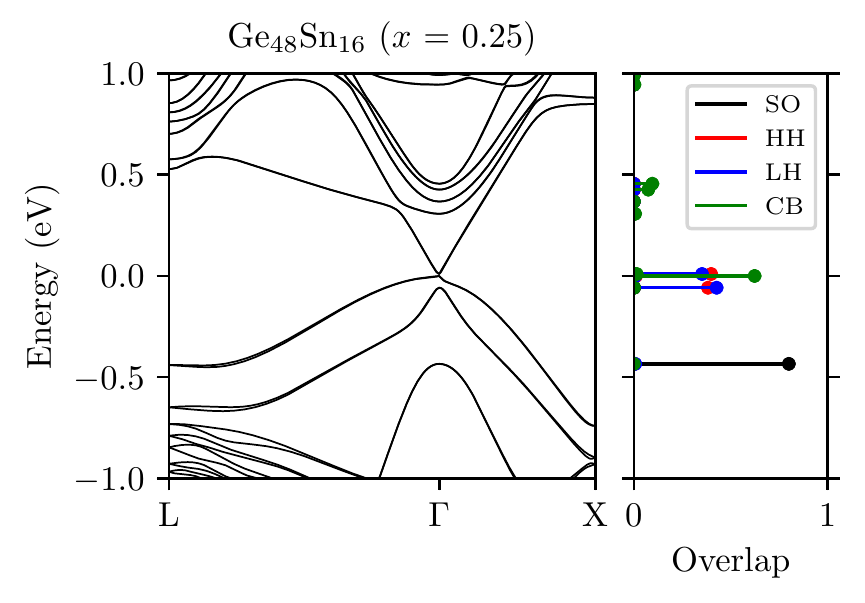}
\caption{Left-hand panel: meta-GGA DFT-calculated band structure of a Ge$_{48}$Sn$_{16}$ ($x = 0.25$) SQS. Right-hand panel: calculated squared overlaps between the alloy SQS zone centre states and the SO ($\Gamma_{7}^{-}$, solid black lines), HH ($\Gamma_{8}^{+}$, solid red lines), LH ($\Gamma_{8}^{+}$, solid blue lines), and CB ($\Gamma_{7}^{-}$, solid green lines) states of a Ge$_{64}$ supercell.}
\label{fig:Ge48Sn16_projections}
\end{figure}

An example of this analysis is presented in Fig.~\ref{fig:Ge48Sn16_projections} for a Ge$_{48}$Sn$_{16}$ ($x = 0.25$) SQS. The left-hand panel in Fig.~\ref{fig:Ge48Sn16_projections} shows the folded supercell band structure. The right-hand panel shows the square of the eigenstate overlaps calculated by projecting the zone-centre CB edge $\Gamma_{7}^{-}$, heavy-hole (HH) and light-hole (LH) VB edge $\Gamma_{8}^{+}$, and spin-orbit split-off (SO) VB edge $\Gamma_{7}^{+}$ states of the unperturbed Ge$_{64}$ host matrix supercell on to the full spectrum of Ge$_{48}$Sn$_{16}$ alloy supercell zone centre states. Here we can clearly identify (i) that the Ge $\Gamma_{7}^{-}$ CB edge character resides primarily on two supercell states, reflecting Sn-induced alloy band mixing which transfers direct Ge $\Gamma_{7}^{-}$ character to the hybridised alloy CB edge, \cite{OHalloran2019} (ii) that the SO band edge is comparatively unperturbed for $x \approx 0.25$, reflecting that Sn incorporation tends primarily to impact the CB structure, \cite{Eckhardt_PRB_2014,Polak2017,OHalloran2019} and (iii) splitting in energy and hybridisation of the LH- and HH-like VB edge states reflecting a loss of cubic symmetry due to short-range alloy disorder and associated local relaxation of the crystal lattice. \cite{Usman_PRB_2013} In the presence of pseudomorphic strain the biaxial component of the strain further reduces the lattice symmetry, acting to push HH-like (LH-like) states upwards (downwards) in energy in the presence of compressive in-plane strain, and vice-versa for tensile strain (cf.~Sec.~\ref{sec:theory_strain}). Despite that short-range alloy disorder drives hybridisation between HH- and LH-like VB states we find generally, as expected for a diamond-like semiconductor, that the state at the alloy Fermi level possesses more Ge HH (LH) $\Gamma_{8}^{+}$ character in the presence of compressive (tensile) in-plane strain (cf.~Eqs.~\eqref{eq:hh_energy_strained} and~\eqref{eq:lh_energy_strained}).

The presence of reduced lattice symmetry and alloy-induced hybridisation complicates interpretation of the alloy electronic structure. Generally, no single alloy supercell state possesses, e.g., 100\% $\Gamma_{7}^{-}$ or $\Gamma_{8}^{+}$ character. \cite{OHalloran2019,Schulz2018} To identify the corresponding energy difference between the two bands $\Gamma_{7}^{-}$-$\Gamma_{8}^{+}$ in our SQS calculations, we select the lowest energy supercell state possessing appreciable Ge $\Gamma_{7}^{-}$ character, and the highest energy supercell state possessing appreciable Ge $\Gamma_{8}^{+}$ character. In the case of semiconducting Ge$_{1-x}$Sn$_{x}$ ($x \lesssim 0.20$) this defines the direct band gap at the Brillouin zone centre, while for higher Sn compositions it defines the maximum inverted (negative) energy gap that must be overcome by quantum confinement to open a band gap in a semimetallic alloy. For reference, calculated supercell band structures and overlaps for relaxed and pseudomorphically strained Ge$_{1-x}$Sn$_{x}$ SQSs are provided as Supplementary Information.


\subsection{Deformation potential theory}
\label{sec:theory_strain}

To aid interpretation of our alloy supercell electronic structure calculations, we apply a deformation potential theory approach. \cite{VandeWalle1989,Krijn_SST_1991} We compare the variation of the strain-dependent alloy supercell $\Gamma_{7}^{-}$ zone-centre CB edge energy, as well as the $\Gamma_{8}^{+}$ HH, $\Gamma_{8}^{+}$ LH, and $\Gamma_{7}^{+}$ SO VB edge energies computed via deformation potential theory to those obtained directly from our TB-mBJ DFT calculations. This comparison also highlights deviations from idealised behaviour associated with alloy band hybridisation effects and short-range alloy disorder. \cite{Polak2017,OHalloran2019,Eales2019}

Pseudomorphic strain in an [001]-oriented cubic crystal results in strain-related shifts to the CB and VB edge energies. The pseudomorphic strain-related shifts to the zone-centre energy of band $n$ = CB, VB are described by a purely diagonal strain tensor and consist of a distinct (i) hydrostatic contribution $\delta E_{\scalebox{0.7}{\text{hy}},n}$, and (ii) biaxial contribution $\delta E_{\scalebox{0.7}{\text{ax}},n}$, which are respectively associated with (i) a change in unit cell volume that is directly proportional to the trace of the strain tensor $\text{Tr} ( \epsilon ) = \epsilon_{xx} + \epsilon_{yy} + \epsilon_{zz}$, and (ii) the biaxial component of the strain tensor $\epsilon_{b} = \epsilon_{zz} - \frac{1}{2} ( \epsilon_{xx} + \epsilon_{yy} )$, which describes the lattice relaxation along the growth direction in response to the in-plane strain imposed via growth on a substrate. Biaxial strain breaks the underlying cubic symmetry of the lattice, lifting the degeneracy of the $p$-like $\Gamma_{8}^{+}$ HH and LH VB edge states. \cite{VandeWalle1989,Krijn_SST_1991} The hydrostatic and biaxial energy shifts are given by $\delta E_{\scalebox{0.7}{\text{hy}},n} = a_{n} \text{Tr} ( \epsilon )$ and $\delta E_{\scalebox{0.7}{\text{ax}},n} = b_{n} \epsilon_{b}$, where $a_{n}$ and $b_{n}$ are, respectively, the hydrostatic and axial deformation potentials associated with band $n$. We denote the CB edge $\Gamma_{7}^{-}$ and VB edge $\Gamma_{8}^{+}$ hydrostatic deformation potentials by $a_{c}$ and $a_{v}$, respectively. The $s$-like $\Gamma_{7}^{-}$ zone-centre CB edge is not impacted by biaxial strain, so that $b_{\scalebox{0.7}{\text{CB}}} = 0$. We denote the (non-zero) VB edge axial deformation potential by $b_{\scalebox{0.7}{\text{VB}}} = b ( \Gamma_{8}^{+} )$ (cf.~Table~\ref{tab:dft_benchmark}).

Taking the zero of energy to lie at the unstrained VB edge, the strain-dependent $\Gamma$-point CB, HH, LH and SO band edge energies in an [001]-oriented pseudomorphic strained layer are given respectively by \cite{VandeWalle1989,Krijn_SST_1991}
\begin{eqnarray}
    E_{\scalebox{0.7}{\text{CB}}} &=& E_{g} + a_{c} \text{Tr} \left( \epsilon \right) \, , \label{eq:cb_energy_strained} \\
    E_{\scalebox{0.7}{\text{HH}}} &=& a_{v} \text{Tr} \left( \epsilon \right) - \delta E_{\scalebox{0.7}{\text{ax,VB}}} \, , \label{eq:hh_energy_strained} \\
    E_{\scalebox{0.7}{\text{LH}}} &=& a_{v} \text{Tr} \left( \epsilon \right) - \frac{1}{2} \left( \Delta_{0} + \delta E_{\scalebox{0.7}{\text{ax,VB}}} \right) \nonumber \\
    &+& \frac{1}{2} \sqrt{ \Delta_{0}^{2} + 2 \Delta_{0} \delta E_{\scalebox{0.7}{\text{ax,VB}}} + 9 \left( \delta E_{\scalebox{0.7}{\text{ax,VB}}} \right)^{2} } \, , \label{eq:lh_energy_strained} \\
    E_{\scalebox{0.7}{\text{SO}}} &=& a_{v} \text{Tr} \left( \epsilon \right) - \frac{1}{2} \left( \Delta_{0} - \delta E_{\scalebox{0.7}{\text{ax,VB}}} \right) \nonumber \\
    &-& \frac{1}{2} \sqrt{ \Delta_{0}^{2} + 2 \Delta_{0} \delta E_{\scalebox{0.7}{\text{ax,VB}}} + 9 \left( \delta E_{\scalebox{0.7}{\text{ax,VB}}} \right)^{2} } \, , \label{eq:so_energy_strained}
\end{eqnarray}

\noindent
where $E_{g} = E ( \Gamma_{7}^{-} ) - E ( \Gamma_{8}^{+} )$ and $\Delta_{0} = E ( \Gamma_{8}^{+} ) - E ( \Gamma_{7}^{+} )$ are, respectively, the direct band gap and VB spin-orbit splitting energy of the unstrained material (cf.~Table~\ref{tab:dft_benchmark}). In the absence of strain, Eqs.~\eqref{eq:cb_energy_strained} --~\eqref{eq:so_energy_strained} yield $E_{\scalebox{0.7}{\text{CB}}} = E_{g}$, $E_{\scalebox{0.7}{\text{HH}}} = E_{\scalebox{0.7}{\text{LH}}} = 0$, and $E_{\scalebox{0.7}{\text{SO}}} = - \Delta_{0}$. In the presence of compressive (tensile) strain, corresponding here to $\epsilon_{xx} < 0$ ($\epsilon_{xx} > 0$), we have $E_{\scalebox{0.7}{\text{HH}}} > E_{\scalebox{0.7}{\text{LH}}}$ ($E_{\scalebox{0.7}{\text{HH}}} < E_{\scalebox{0.7}{\text{LH}}}$).

Choosing the $z$-axis to align to the growth direction, for an [001]-oriented pseudomorphic layer we have $\epsilon_{xx} = \epsilon_{yy} = \frac{ a_{\scalebox{0.5}{\text{S}}} - a_{\scalebox{0.5}{\text{L}}} }{ a_{\scalebox{0.5}{\text{L}}} }$ as the fractional lattice mismatch in the plane perpendicular to the growth direction due to lattice mismatch between the layer having lattice constant $a_{\scalebox{0.5}{\text{L}}}$, and the substrate having lattice constant $a_{\scalebox{0.5}{\text{S}}}$. Similarly, $\epsilon_{zz} = - \frac{ 2 C_{12} }{ C_{11} } \epsilon_{xx}$ describes the corresponding relaxation of the layer along the growth direction in response to the in-plane strain. \cite{VandeWalle1989,Krijn_SST_1991} As such, $\text{Tr} ( \epsilon ) = 2 ( 1 - \frac{ C_{12} }{ C_{11} } ) \epsilon_{xx}$ and $\epsilon_{b} = - ( 1 + \frac{ 2 C_{12} }{ C_{11} } ) \epsilon_{xx}$. Under hydrostatic strain, $\epsilon_{xx} = \epsilon_{yy} = \epsilon_{zz}$, the strain-induced change in band gap is given via Eqs.~\eqref{eq:cb_energy_strained} and~\eqref{eq:hh_energy_strained} as $a_{g} \text{Tr} ( \epsilon )$, where $a_{g} = a_{c} - a_{v}$ is the direct-gap hydrostatic deformation potential (i.e.~$a_{g} ( \Gamma_{7}^{-}$-$\Gamma_{8}^{+} )$; cf.~Table~\ref{tab:dft_benchmark}), related to the band gap pressure coefficient as $\frac{ d E_{g} }{ dP } = - \frac{ a_{g} }{ B }$, where $B = \frac{1}{3} ( C_{11} + 2 C_{12} )$ is the bulk modulus. \cite{Wei1999}

To evaluate Eqs.~\eqref{eq:cb_energy_strained} --~\eqref{eq:so_energy_strained} we have computed the elastic constants and CB and VB edge deformation potentials for Ge and $\alpha$-Sn directly via TB-mBJ DFT (cf.~Table~\ref{tab:dft_benchmark}). The hydrostatic band gap deformation potentials $a_{g}$ were obtained by calculating the direct and indirect energy gaps $\Gamma_{7}^{-}$-$\Gamma_{8}^{+}$ and L$_{6}^{+}$-$\Gamma_{8}^{+}$ for Ge and $\alpha$-Sn under applied hydrostatic pressure. The VB edge axial deformation potentials $b$ were obtained by calculating the splitting between the HH and LH VB edge energies under an applied volume-preserving (to first order in $\epsilon$) uniaxial strain along [001]. To analyse our Ge$_{1-x}$Sn$_{x}$ alloy SQS calculations, we linearly interpolate the deformation potentials of Ge and $\alpha$-Sn to obtain deformation potentials for Ge$_{1-x}$Sn$_{x}$, and also consider best-fit deformation potentials which reflect alloy-induced hybridisation between distinct host matrix eigenstates (cf.~Sec.~\ref{sec:results_transition}). The elastic constants used in our calculations were computed via LDA DFT for Ge and $\alpha$-Sn via stress tensors computed under applied high-symmetry strain branches. \cite{Caro_JPCM_2012,Tanner_submitted_2020} For completeness, we list also in Table~\ref{tab:dft_benchmark} the elastic constant $C_{44}$ and VB edge shear deformation potential $d$, which do not enter into the present analysis. Repeating the elastic constant calculations for the Ge$_{1-x}$Sn$_{x}$ alloy SQSs, we find significant bowing of the elastic constants. The bowing parameters for $C_{11}$, $C_{12}$ and $C_{44}$ were respectively computed as 40.87, 4.82 and 30.41 GPa, and were used to interpolate the elastic constants used in our calculation of the components of the strain tensor via Eqs.~\eqref{eq:cb_energy_strained} --~\eqref{eq:so_energy_strained}, and also in our critical thickness calculations below.


\subsection{Critical thickness}
\label{sec:theory_critical}

The growth of pseudomorphically strained layers proceeds for layer thicknesses $t \lesssim t_{c}$, beyond which plastic relaxation via the formation of dislocations and related crystalline defects becomes energetically favourable. \cite{OReilly1989} For a pseudomorphically strained layer this critical thickness $t_{c}$ can be computed as the root of the intrinsic equation \cite{Voisin1988}

\begin{equation}
     t_{c} = \frac{ a_{\scalebox{0.7}{\textrm{S}}} }{ 8 \sqrt{2} \pi |\epsilon_{xx}| } \left( \frac{ 4 - \sigma }{ 1 + \sigma } \right) \left( 1 + \ln \left( \frac{ \sqrt{2} t_{c} }{ a_{\scalebox{0.7}{\textrm{S}}} } \right) \right) \, ,
     \label{eq:critical_thickness}
\end{equation}

\noindent
where $a_{\scalebox{0.7}{\textrm{S}}}$ is the substrate lattice constant, $\epsilon_{xx}$ is the lattice mismatch between the substrate and the layer in the plane perpendicular to the [001] direction, and $\sigma = \frac{ C_{12} }{ C_{11} + C_{12} }$ is Poisson's ratio for the layer. While there exist numerous approaches to compute $t_{c}$, \cite{OReilly1989} Eq.~\eqref{eq:critical_thickness} has been demonstrated to produce estimates which are in good agreement with experimental measurements for a range of semiconductor materials, including group-IV Si$_{x}$Ge$_{1-x}$ and III-V In$_{y}$Ga$_{1-y}$As alloys, \cite{Voisin1988,OReilly1989} as well as highly-mismatched III-V alloys containing nitrogen or bismuth. \cite{Tomic2003,Broderick2017} As we describe in Sec.~\ref{sec:results_transition} below, the estimates of $t_{c}$ computed via Eq.~\eqref{eq:critical_thickness} are in quantitative agreement with the results of LDA DFT structural relaxation for disordered Ge$_{1-x}$Sn$_{x}$ alloy supercells.


\section{Results}
\label{sec:results}

We next consider the results of our calculations, and analysis of the band gap and semimetal to semiconductor transition in Ge$_{1-x}$Sn$_{x}$. In Secs.~\ref{sec:GeSn_unstrained} and~\ref{sec:results_strained} we analyse the electronic structure of relaxed and pseudomorphically strained alloys. In Sec.~\ref{sec:results_transition}, we compare the results of our strained alloy supercell DFT calculations to the deformation potential theory of Sec.~\ref{sec:theory_strain}, analyse critical thickness limitations associated with growth of Ge$_{1-x}$Sn$_{x}$ on Ge, ZnTe and CdTe substrates, and explore the effects of tensile strain to reduce the Sn composition required to close the band gap and achieve a semimetallic band structure in pseudomorphic Ge$_{1-x}$Sn$_{x}$.


\subsection{Band gap evolution in relaxed Ge$_{1-x}$Sn$_{x}$ alloys}
\label{sec:GeSn_unstrained}

\begin{figure}
\centering
\includegraphics[width=0.94\columnwidth]{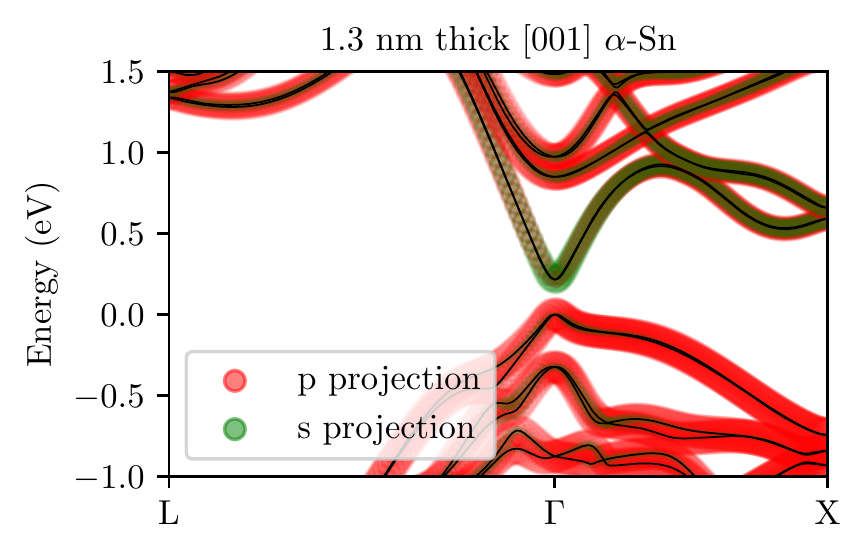}
\caption{Opening of a direct band gap via quantum confinement in a semimetallic $\alpha$-Sn thin film: $s$ (red) and $p$ (green) orbital resolved band structure of an (001)-oriented, H-terminated film of thickness 1.3 nm calculated via LDA-DFT.}
\label{fig:1nm_sn_pdos}
\end{figure}

Ge possesses an indirect fundamental band gap between the L$_{6}^{+}$ CB minimum and $\Gamma_{8}^{+}$ VB maximum, while $\alpha$-Sn possesses a direct inverted band gap between the $\Gamma_{7}^{-}$ and $\Gamma_{8}^{+}$ zone-centre eigenstates. Being an inverted-gap semimetal, $\alpha$-Sn has no band gap above the $\Gamma_{8}^{+}$ VB edge, but does display a vanishing DOS where the VB and CB meet (i.e.~at the Fermi energy). It has been demonstrated theoretically that quantum confinement effects can be exploited to open a direct band gap in $\alpha$-Sn thin films or nanowires, by pushing the $s$-like $\Gamma_{7}^{-}$ states higher in energy than the $p$-like $\Gamma_{8}^{+}$ states. \cite{Kuefner2013} As an example of this behaviour, we present in Fig.~\ref{fig:1nm_sn_pdos} the LDA DFT-calculated orbital resolved band structure of a 1.3 nm thick, [001]-oriented and H-terminated $\alpha$-Sn thin film. The black solid lines show the calculated energy bands, demonstrating the opening of a direct band gap above the Fermi level, chosen in Fig.~\ref{fig:1nm_sn_pdos} to lie at the zero of energy. The weighted green and red colour of the overlaid circles respectively denote the total $s$ and $p$ orbital character of the associated band states. Examining the orbital character, we note that the zone-centre CB minimum is purely $s$-like, reflecting that it originates from an upward shift in energy of the $\Gamma_{7}^{-}$ states of $\alpha$-Sn in response to quantum confinement. We note that the TB-mBJ exchange-correlation potential applied in our analysis of bulk-like Ge$_{1-x}$Sn$_{x}$ supercells is not applicable in thin film calculations, as the potential becomes singular in the presence of vanishing charge density. \cite{Tran2009} The LDA-calculated 219 meV band gap in Fig.~\ref{fig:1nm_sn_pdos} is therefore likely underestimated, \cite{Sanchez-Soares2016a} but nonetheless, as a known underestimate of the band gap energy, strongly suggests that a direct band gap can be opened in $\alpha$-Sn thin films via quantum confinement.

\begin{figure}
\centering
\includegraphics[width=0.94\columnwidth]{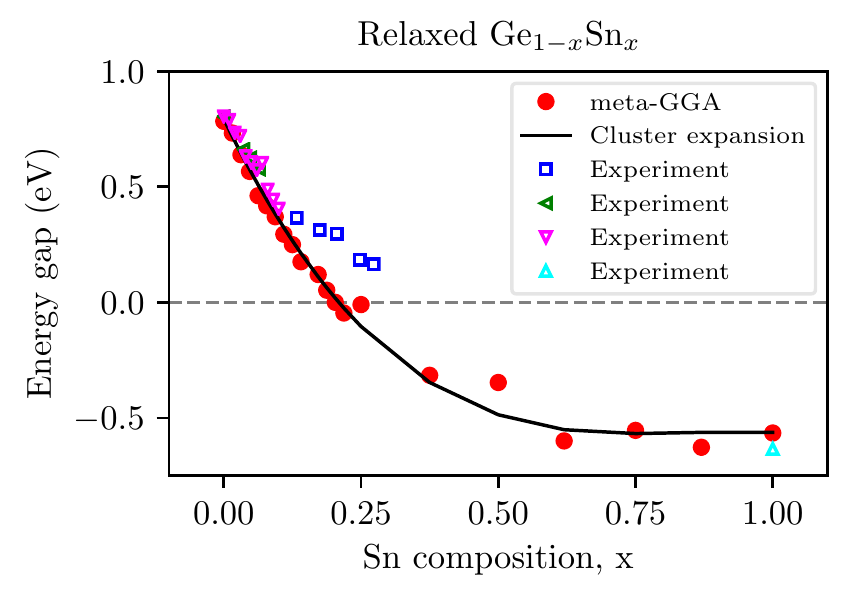}
\caption{Meta-GGA DFT-calculated band gap as a function of Sn composition $x$ of Ge$_{1-x}$Sn$_{x}$ alloy SQSs across the full alloy composition range. The solid black line represents the corresponding cluster expansion calculation. Closed green triangles, pink triangles and blue squares denote measured low temperature band gaps in the semiconducting regime, \cite{Lin2012,Tran_JAP_2016,MenendezA} and the light blue triangle denotes the experimentally measured inverted gap of $\alpha$-Sn\cite{Booth1968_a}. The horizontal dashed line denotes zero band gap, demarcating between semiconducting and semimetallic band structure.}
\label{fig:GeSn_band_gap_unstrained}
\end{figure}

Previous theoretical calculations have predicted the evolution and nature of the Ge$_{1-x}$Sn$_{x}$ alloy band gap across the full composition range, from Ge to $\alpha$-Sn. \cite{MoontragoonIkonicHarrison2007,Polak2017,Lan2017} Beginning with the indirect-gap semiconductor Ge, Sn incorporation drives rapid band gap reduction and an indirect to direct-gap transition for $x \lesssim 0.10$. Beyond this composition, the direct band gap decreases in magnitude until it closes between $0.2 \lesssim x \lesssim 0.4$, at which point the alloy becomes a zero-gap semimetal. For higher Sn compositions the alloy band structure is predicted to be semimetallic and possesses an inverted $\Gamma_{7}^{-}$-$\Gamma_{8}^{+}$ band ordering. We have firstly employed TB-mBJ DFT calculations to verify the evolution of the Ge$_{1-x}$Sn$_{x}$ band gap using relaxed 64-atom SQSs. The results of these calculations are summarised in Fig.~\ref{fig:GeSn_band_gap_unstrained}, where closed red circles denote the calculated fundamental alloy band gap. The solid black line shows the corresponding fundamental band gap calculated via a cluster expansion approach, \cite{Sanchez1984} allowing for prediction of alloy properties as the SQS lattice correlation functions approximate those of a random alloy. We note that the cluster expansion approach has been recently applied to quantify the formation enthalpy and alloy miscibility in Ge$_{1-x}$Sn$_{x}$. \cite{Miscibility} The dashed black line in Fig.~\ref{fig:GeSn_band_gap_unstrained} denotes the composition at which a zero band gap is predicted to occur thereby demarcating the Sn compositions at which the alloy is either semiconducting or semimetallic. Negative values refer to the inverted (semimetallic) $\Gamma_{7}^{-}$-$\Gamma_{8}^{+}$ energy gaps.


At low Sn compositions $x \lesssim 0.20$, we calculate a rapid band gap reduction with increasing $x$ (the indirect- to direct-gap transition occurring in this composition range is discussed in Sec.~\ref{sec:results_transition}). Our calculated band gaps are also compared to experimental measurements in Fig.~\ref{fig:GeSn_band_gap_unstrained}. Generally, Ge$_{1-x}$Sn$_{x}$ is grown as a thin strained, or thick relaxed, epitaxial layer on a substrate having a lattice constant close to that of Ge. The resultant pseudomorphic strain complicates the comparison between calculated and measured band gaps, due to the impact of strain on the magnitude of the band gap (cf.~Sec.~\ref{sec:theory_strain}). In Fig.~\ref{fig:GeSn_band_gap_unstrained}, we therefore compare the results of our TB-mBJ DFT calculations to experimental data in which the impact of pseudomorphic strain is minimised. Specifically, we compare our calculations to the results of photo-modulated reflectance (Ref.~\onlinecite{Lin2012}; open green triangles) and optical absorption (Refs.~\onlinecite{Tran_JAP_2016} and~\onlinecite{MenendezA}; open pink triangles and blue squares, respectively) spectroscopic measurements. The photo-modulated reflectance measurements of Ref.~\onlinecite{Lin2012} were performed on epitaxial layers grown on nominally lattice-matched In$_{y}$Ga$_{1-y}$As buffer layers, thereby minimising strain. The absorption measurements of Ref.~\onlinecite{Tran_JAP_2016} were performed on strained Ge$_{1-x}$Sn$_{x}$ epitaxial layers, but the data were then corrected to account for the strain-induced effects on the band gap, based on strain values extracted from x-ray diffraction measurements. The absorption measurements of Ref.~\onlinecite{MenendezA} were performed on thick, relaxed epitaxial layers grown on Si substrates. We note excellent quantitative agreement between our calculated band gap as a function of Sn composition $x$ and that measured by Lin et al. \cite{Lin2012} and by Tran et al. \cite{Tran_JAP_2016} However, our calculations deviate from the measurements of Xu et al \cite{MenendezA} for $0.15 \lesssim x \lesssim 0.25$. Given the large lattice mismatch between Ge$_{1-x}$Sn$_{x}$ and the Si substrates employed by Xu et al, it is possible that there remains residual compressive strain associated with partial relaxation in their samples which would act to increase the band gap, possibly explaining why our calculations underestimate those measured band gaps. Finally, the open light-blue triangle represents the $\alpha$-Sn inverted direct $\Gamma_{7}^{-}$-$\Gamma_{8}^{+}$ energy gap, measured by Booth and Ewald \cite{Booth1968_a} via magnetoresistance (cf.~Table~\ref{tab:dft_benchmark}). Overall, we note good quantitative agreement between our theoretical calculations and comparable experimental data.


We predict a closing of the band gap in relaxed Ge$_{1-x}$Sn$_{x}$ for $x \approx 0.21$. Analysis of the supercell electronic structure for SQSs having $x \gtrsim 0.21$ reveals that the alloy is semimetallic, in agreement with recent theoretical calculations. \cite{Polak2017,Lan2017} For Sn compositions $x \gtrsim 0.50$, we predict that the magnitude of the inverted $\Gamma_{7}^{-}$-$\Gamma_{8}^{+}$ energy gap remains close to the value $-0.59$ eV calculated for $\alpha$-Sn (cf.~Table~\ref{tab:dft_benchmark}). There are, however, differing predictions in the literature regarding the precise nature of the semimetallic alloy band structure for $x \gtrsim 0.20$. Using empirical pseudopotential calculations in the virtual crystal approximation, Lan et al. \cite{Lan2017} predicted that the band ordering remains topologically non-trivial in this composition range with $\Gamma_{7}^{-}$ states lying lower and higher in energy than $\Gamma_{8}^{+}$ and $\Gamma_{7}^{+}$ states, respectively. Conversely, by applying zone unfolding to DFT calculations for SQS supercells, Polak et al \cite{Polak2017} predicted the presence of a topologically trivial ``inverted SO'' band structure for $0.45 \lesssim x \lesssim 0.85$, in which the $\Gamma_{7}^{-}$ states drop lower in energy than the $\Gamma_{7}^{+}$ states. However, it is noteworthy that the quadratic composition dependent band gap fit applied by Polak et al to extract this conclusion is poorer in quality for $x \gtrsim 0.30$ compared to its accuracy at lower Sn compositions.

To resolve these contrasting predictions, we have undertaken a direct analysis of the alloy supercell zone-centre states, via projection onto reference $\Gamma_{7}^{-}$, $\Gamma_{8}^{+}$ and $\Gamma_{7}^{+}$ states calculated for a Ge$_{64}$ supercell (cf.~Fig.~\ref{fig:Ge48Sn16_projections}). Applying this quantitative approach, we find that the alloy state retaining the largest calculated $\Gamma_{7}^{-}$ character reduces in energy with increasing $x$, approaching the state retaining the largest calculated $\Gamma_{7}^{+}$ character. At $x = 0.625$ (Ge$_{24}$Sn$_{40}$ SQS), we calculate that these Ge $\Gamma_{7}^{-}$- and $\Gamma_{7}^{+}$-derived alloy states become quasi-degenerate, and remain so for Sn compositions up to $x = 0.875$ (corresponding to a Ge$_{8}$Sn$_{56}$ SQS), beyond which composition the relative energy of the Ge $\Gamma_{7}^{-}$-derived alloy state increases to reach its value in $\alpha$-Sn -- i.e.~lying $\approx 100$ meV above $\Gamma_{7}^{+}$ in energy (cf.~Table~\ref{tab:dft_benchmark}). In only one of the supercells analysed was a Ge $\Gamma_{7}^{-}$-derived alloy state found to lie lower in energy than a $\Gamma_{7}^{+}$-derived state, and in that case only by 10 meV and with additional Ge $\Gamma_{7}^{-}$ character residing on higher energy alloy states. Our calculations therefore suggest that semimetallic Ge$_{1-x}$Sn$_{x}$ most likely retains topologically non-trivial band ordering for $0.21 \lesssim x \leq 1$, but we note that the specific band ordering close to $x = 0.60$ for small supercell calculations is determined in part by the impact of alloy-induced band hybridisation, arising due to the precise reduction in symmetry associated with the specific alloy disorder and lattice relaxation present in a given alloy SQS.

We attribute deviations between previous predictions to the simplified manner in which the alloy band structure was analysed, generally based on either (i) assuming virtual crystal-like behaviour and simple polynomial fits to composition-dependent band gaps obtained from theoretical data, or (ii) extrapolation of low $x$ experimental data across the full composition range. In particular, large uncertainties associated with extrapolation of experimental data have produced divergent predictions, \cite{MenendezA} including (i) that the magnitude of the inverted $\Gamma_{7}^{-}$-$\Gamma_{8}^{+}$ alloy energy gap exceeds that of $\alpha$-Sn by up to 0.2 eV, or (ii) that the direct $\Gamma_{7}^{-}$-$\Gamma_{8}^{+}$ band gap remains open with a constant magnitude $\approx 0.1$ eV up to $x \approx 0.70$. We emphasise that our conclusions here are based on direct investigation of the character and evolution of hybridised alloy supercell eigenstates calculated from first principles.


Our calculations indicate that the magnitude of the direct inverted band gap in relaxed Ge$_{1-x}$Sn$_{x}$ can be tuned between zero and close to that of $\alpha$-Sn for $0.21 \lesssim x \lesssim 0.50$. This provides the ability to reduce the magnitude of the inverted $\Gamma_{7}^{-}$-$\Gamma_{8}^{+}$ energy gap that must be overcome to open a band gap via quantum confinement. Alloying therefore provides a degree of freedom in addition to nanostructuring, which can be utilised to engineer a semimetal-to-semiconductor transition via quantum confinement. In what follows, we turn our attention to the band structure of pseudomorphically strained Ge$_{1-x}$Sn$_{x}$ alloys, on which basis we then quantify the impact of alloying and strain in epitaxial thin films, in order to provide further control over the inverted energy gap and hence the semimetal-to-semiconductor transition.


\subsection{Band structure evolution in pseudomorphically strained Ge$_{1-x}$Sn$_{x}$}
\label{sec:results_strained}

Having analysed the nature and evolution of the Ge$_{1-x}$Sn$_{x}$ band gap for the full composition range in relaxed alloys, we now consider pseudomorphically strained Ge$_{1-x}$Sn$_{x}$ grown on (i) Ge, (ii) ZnTe, and (iii) CdTe substrates. ZnTe and CdTe have lattice constants close to those calculated for zb-GeSn \cite{OHalloran2019,Tanner_submitted_2020} and $\alpha$-Sn, respectively. For each substrate, we take the SQSs of Sec.~\ref{sec:GeSn_unstrained} and simulate [001]-oriented pseudomorphically strained Ge$_{1-x}$Sn$_{x}$ by fixing the supercell lattice constants along the [100] and [010] directions to mimic the substrate lattice constant. Next, we perform a LDA relaxation of the [001] lattice supercell parameter and all ionic positions. We then perform a TB-mBJ electronic structure calculation for these pseudmorphically strained SQSs, and again quantify the character of the alloy supercell eigenstates by projecting them onto reference states from a Ge$_{64}$ supercell. On this basis we assign a $\Gamma_{7}^{-}$-$\Gamma_{8}^{+}$ energy gap for each strained SQS, as above.

\begin{figure}
\centering
\includegraphics[width=0.94\columnwidth]{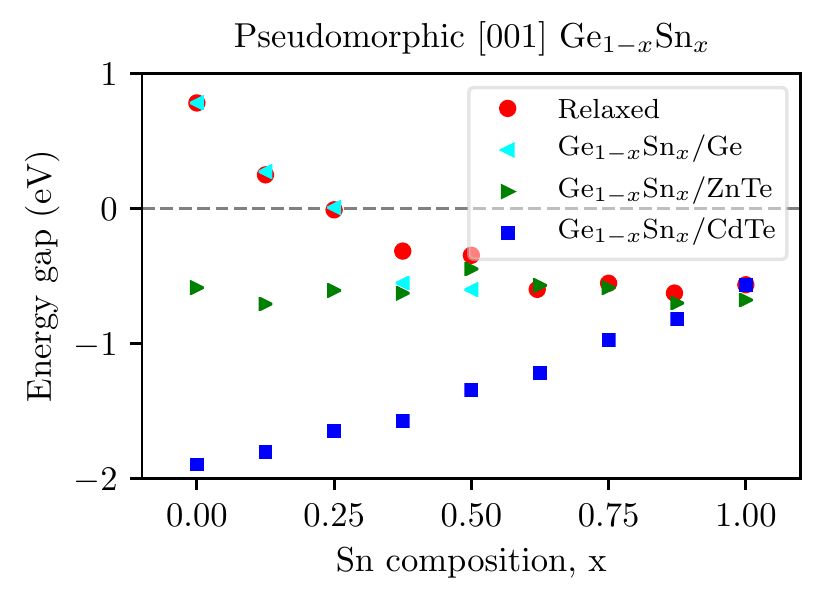}
\caption{Meta-GGA DFT-calculated band gap as a function of Sn composition $x$ of Ge$_{1-x}$Sn$_{x}$ alloy SQSs which are freely relaxed (closed red circles), or pseudomorphically strained and internally relaxed to correspond to growth on Ge (closed light blue triangles), ZnTe (closed green triangles), or CdTe (closed blue squares) substrates. The horizontal dashed line denotes zero band gap, demarcating between semiconducting and semimetallic band structure.}
\label{fig:GeSn_band_gap_strained}
\end{figure}

The results of these calculations are summarised in Fig.~\ref{fig:GeSn_band_gap_strained}, where the closed red circles repeat the results of the unstrained calculations of Fig.~\ref{fig:GeSn_band_gap_unstrained} for reference. The closed light blue triangles, closed green triangles and closed blue squares in Fig.~\ref{fig:GeSn_band_gap_strained} respectively show the calculated variation of the $\Gamma_{7}^{-}$-$\Gamma_{8}^{+}$ energy difference (direct band gap) across the full composition range for Ge$_{1-x}$Sn$_{x}$ alloys pseudomorphically strained to [001]-oriented Ge, ZnTe and CdTe substrates. Our LDA total energy calculations suggest that, when grown on Ge, it is energetically favourable for Ge$_{1-x}$Sn$_{x}$ to become amorphous for Sn compositions $x \gtrsim 0.56$. \cite{Miscibility} Interpolating the LDA-calculated Ge and $\alpha$-Sn lattice constants of Table~\ref{tab:dft_benchmark} using a bowing parameter of 0.056, \cite{Miscibility} $x = 0.56$\% corresponds to a (compressive) lattice mismatch $\epsilon_{xx} \approx -7.8$\%. As such, the associated Ge$_{1-x}$Sn$_{x}$/Ge band gap is shown only up to $x=0.5$ in Fig.~\ref{fig:GeSn_band_gap_strained} (light blue triangles). As we will describe in Sec.~\ref{sec:results_transition}, this conclusion is in quantitative agreement with our estimation of the Ge$_{1-x}$Sn$_{x}$/Ge critical thickness.

Comparing the band gaps of relaxed Ge$_{1-x}$Sn$_{x}$ and compressively strained Ge$_{1-x}$Sn$_{x}$/Ge in Fig.~\ref{fig:GeSn_band_gap_strained}, we note that the latter closely track the former for $x \lesssim 0.30$. The strained Ge$_{1-x}$Sn$_{x}$/Ge band gaps are calculated to slightly exceed those of relaxed Ge$_{1-x}$Sn$_{x}$ in this composition range, reflecting that compressive pseudomorphic strain acts to slightly increase the band gap (cf.~Eqs.~\eqref{eq:cb_energy_strained} --~\eqref{eq:lh_energy_strained}). As the Sn composition is increased above $x= 0.30$, we note that the calculated magnitude of the inverted (negative) $\Gamma_{7}^{-}$-$\Gamma_{8}^{+}$ energy gap of strained Ge$_{1-x}$Sn$_{x}$/Ge exceeds that in relaxed Ge$_{1-x}$Sn$_{x}$, i.e.~the Ge $\Gamma_{7}^{-}$-derived alloy state is lower in energy relative to the Ge $\Gamma_{8}^{+}$-derived state that corresponds to the alloy VB edge. This is contrary to the expected trend based on the compressive strain in the Ge$_{1-x}$Sn$_{x}$/Ge SQSs within this composition range. Based on our analysis of the character of the eigenstates for the corresponding SQSs (cf.~Supplementary Information), we attribute this behaviour to alloy-induced hybridisation which distributes the Ge $\Gamma_{7}^{-}$ character over an energy range $\lesssim 1$ eV in width. If instead of assigning as $\Gamma_{7}^{-}$ the lowest energy alloy state possessing appreciable Ge $\Gamma_{7}^{-}$, as described in Sec.~\ref{sec:theory_sqs}, the weighted average energy of the supercell eigenstates is instead calculated -- using the computed Ge $\Gamma_{7}^{-}$ character of each state as the weight for that state's energy -- this trend is reversed, and leads to Ge$_{1-x}$Sn$_{x}$/Ge inverted band gaps which are smaller in magnitude than in relaxed Ge$_{1-x}$Sn$_{x}$, as expected for compressive strain.


Next, we note that the inverted band gap in tensile strained Ge$_{1-x}$Sn$_{x}$/CdTe, shown using closed blue squares in Fig.~\ref{fig:GeSn_band_gap_strained}, increases in magnitude as Ge is incorporated into $\alpha$-Sn. This behaviour is in line with the expected reduction in energy of $\Gamma_{7}^{-}$ states relative to $\Gamma_{8}^{+}$ states in the presence of tensile in-plane strain (cf.~Eqs.~\eqref{eq:cb_energy_strained} and~\eqref{eq:lh_energy_strained}). Our calculated energy gaps for Ge$_{1-x}$Sn$_{x}$/CdTe therefore suggest that growth of tensile strained Ge$_{1-x}$Sn$_{x}$ will increase the magnitude of the inverted $\Gamma_{7}^{-}$-$\Gamma_{8}^{+}$ energy gap, making it challenging to for an energy band gap to be opened via quantum confinement in a nanostructure.

Considering Ge$_{1-x}$Sn$_{x}$/ZnTe (closed green triangles in Fig.~\ref{fig:GeSn_band_gap_strained}), the substrate introduces compressive (tensile) in-plane strain for $x \geq 0.54$ ($x \leq 0.54$). In both cases, we note that the magnitude of the inverted $\Gamma_{7}^{-}$-$\Gamma_{8}^{+}$ energy gap remains approximately constant independent of changes in Sn composition. Reducing the Sn composition from $x = 0.54$ results in tensile in-plane strain. In this composition range, a reduction in $x$ acts to increase the energy of $\Gamma_{7}^{-}$-like states relative to $\Gamma_{8}^{+}$-like states, resulting in a reduction of the magnitude of the inverted band gap. Due to the strain present in pseudomorphic Ge$_{1-x}$Sn$_{x}$/ZnTe, the composition-dependent band edge energy shifts are counteracted by the tensile in-plane strain which acts to decrease (increase) the energy of $\Gamma_{7}^{-}$-like (LH $\Gamma_{8}^{+}$-like) states. This competition between Sn composition and strain is mirrored in the compressively strained regime, leading to a Ge$_{1-x}$Sn$_{x}$/ZnTe inverted energy gap which remains relatively constant across the full composition range. We note that the calculated 448 meV magnitude of the inverted band gap in approximately lattice-matched Ge$_{0.5}$Sn$_{0.5}$ (Ge$_{32}$Sn$_{32}$ SQS) is approximately two-thirds of the magnitude of the inverted $\Gamma_{7}^{-}$-$\Gamma_{8}^{+}$ energy gap of $\alpha$-Sn. An equivalent TB-mBJ band structure calculation predicts that zb-GeSn (i.e.~an ordered Ge$_{0.5}$Sn$_{0.5}$ alloy) is a zero-gap semiconductor, \cite{OHalloran2019} with the significant difference in band gap between ordered zb-GeSn and disordered Ge$_{0.5}$Sn$_{0.5}$ highlighting that, at fixed Sn composition, the magnitude of the inverted energy gap is strongly influenced by short-range alloy disorder. \cite{OHalloran2019}

The magnitude of the inverted band gap shown in Fig.~\ref{fig:GeSn_band_gap_strained} for Ge$_{1-x}$Sn$_{x}$/ZnTe is (i) associated with maximally random alloy disorder due to the use of SQSs in our calculations, and (ii) explicitly assigned the largest possible value, via our previously described explicit assignment of the lowest energy alloy state possessing appreciable Ge $\Gamma_{7}^{-}$ character as ``$\Gamma_{7}^{-}$'' in our energy gap calculations (cf.~Sec.~\ref{sec:theory_sqs}). As such, the magnitude of the inverted band gap in Ge$_{1-x}$Sn$_{x}$/ZnTe indicated in Fig.~\ref{fig:GeSn_band_gap_strained} should be considered an upper limit, which is likely reduced both in a real alloy (where more states are free to hybridise than in the small supercell calculations considered here, leading to less energetic spread of Ge $\Gamma_{7}^{-}$ character), and in the presence of partial alloy ordering (where the inverted band gap is expected to move towards the zero band gap of ordered zb-GeSn). In addition to alloying, strain is seen to provide another degree of freedom that can be exploited to engineer the electronic structure in Ge$_{1-x}$Sn$_{x}$ to be either semimetallic or semiconducting. Choosing a substrate and alloy composition determines the magnitude of the inverted $\Gamma_{7}^{-}$-$\Gamma_{8}^{+}$ energy gap that needs to be overcome to achieve a semimetal-to-semiconductor transition driven by quantum confinement.


\subsection{Engineering the electronic structure of pseudomorphic Ge$_{1-x}$Sn$_{x}$}
\label{sec:results_transition}

We next consider the impact of strain on the band gap using deformation potential theory, and describe trends in light of the results of our DFT alloy SQS calculations. This provides an efficient means by which to predict strain-dependent energy gaps, while also providing qualitative insight into alloy-induced band mixing effects. We then predict critical thickness limitations for Ge$_{1-x}$Sn$_{x}$ alloys grown on Ge, ZnTe and CdTe substrates. Finally, we describe the alloy composition and strain for determining whether a given bulk alloy will be semiconducting or semimetallic, and provide general recommendations for the growth of pseudomorphically strained Ge$_{1-x}$Sn$_{x}$ thin films in which the band gap is closed within known alloy miscibility restrictions -- i.e.~at the lowest possible Sn composition, and within critical thickness limits.

\begin{figure}
\centering
\includegraphics[width=0.94\columnwidth]{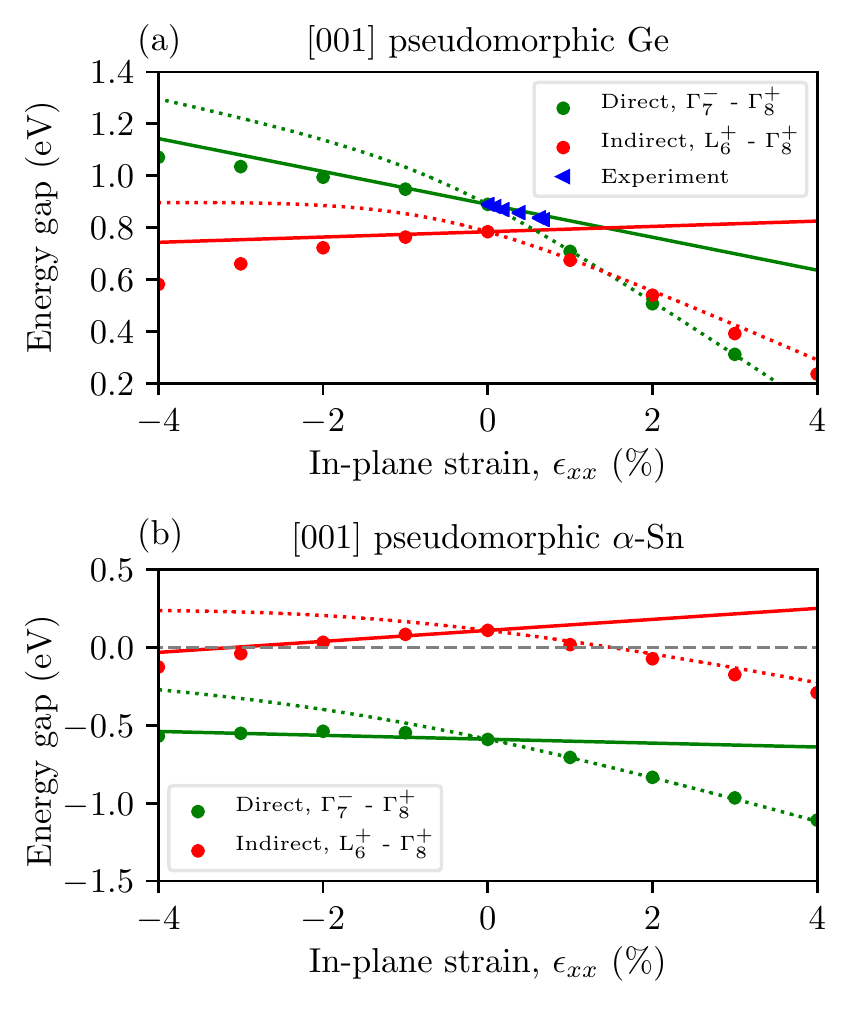}
\caption{(a) Meta-GGA DFT-calculated indirect (L$_{6}^{+}$-$\Gamma_{8}^{+}$; closed red circles) and direct ($\Gamma_{7}^{-}$-$\Gamma_{8}^{+}$; closed green circles) band gaps of pseudmorphically strained (a) Ge, and (b) $\alpha$-Sn, as a function of in-plane strain $\epsilon_{xx}$. Solid and dashed lines show band gaps calculated via deformation potential theory, using Eqs.~\eqref{eq:cb_energy_strained} --~\eqref{eq:so_energy_strained} in conjunction with the data of Table~\ref{tab:dft_benchmark}. Solid (dashed) green lines show the direct band gap between $\Gamma_{7}^{-}$ CB and HH- (LH-) like $\Gamma_{8}^{+}$ VB states. Solid (dashed) red lines show the indirect band gap between L$_{6}^{+}$ CB and HH- (LH-) like $\Gamma_{8}^{+}$ VB states. Closed blue triangles show the experimental measurements of Ref.~\onlinecite{ElKurdi2010}.}
\label{fig:Ge_and_Sn_strained}
\end{figure}

We begin our deformation potential theory analysis by considering the band gap of pseudomorphically strained Ge and $\alpha$-Sn. The results of these calculations are summarised in Figs.~\ref{fig:Ge_and_Sn_strained}(a) and~\ref{fig:Ge_and_Sn_strained}(b), respectively. In each case, closed green circles show the TB-mBJ DFT-calculated direct $\Gamma_{7}^{-}$-$\Gamma_{8}^{+}$ energy gap, as a function of in-plane strain $\epsilon_{xx}$. We note that in-plane strain values $< 0$ ($> 0$) correspond to compressive (tensile) strain (cf.~Sec.~\ref{sec:theory_strain}). Closed red circles in Figs.~\ref{fig:Ge_and_Sn_strained}(a) and~\ref{fig:Ge_and_Sn_strained}(b) show the TB-mBJ DFT-calculated indirect L$_{6}^{+}$-$\Gamma_{8}^{+}$ energy gap. Green and red lines in Figs.~\ref{fig:Ge_and_Sn_strained}(a) and~\ref{fig:Ge_and_Sn_strained}(b) respectively show the corresponding energy gaps calculated via Eqs.~\eqref{eq:cb_energy_strained} --~\eqref{eq:lh_energy_strained}, using the parameters listed in Table~\ref{tab:dft_benchmark}. Solid and dashed lines respectively show the band gaps calculated with respect to HH- and LH-like $\Gamma_{8}^{+}$ VB states, where we recall that $E_{\scalebox{0.7}{\text{HH}}} > E_{\scalebox{0.7}{\text{LH}}}$ for $\epsilon_{xx} < 0$ (and vice versa), so that solid (dashed) lines represent the direct $\Gamma_{7}^{-}$-$\Gamma_{8}^{+}$ and indirect L$_{6}^{+}$-$\Gamma_{8}^{+}$ band gaps associated with HH- (LH-) like $\Gamma_{8}^{+}$ VB edge states in the compressive (tensile) strained regime. Closed blue triangles in Fig.~\ref{fig:Ge_and_Sn_strained}(a) show experimental measurements of the direct band gap of tensile-strained Ge epitaxial layers. \cite{ElKurdi2010} We note in Figs.~\ref{fig:Ge_and_Sn_strained}(a) and~\ref{fig:Ge_and_Sn_strained}(b) good quantitative agreement between the full DFT (circles) and deformation potential theory (lines) calculations. We attribute increased deviation between the DFT-calculated and deformation potential theory results at high compressive strain in Ge to increasing non-linear strain contributions, which are not captured by Eqs.~\eqref{eq:cb_energy_strained} --~\eqref{eq:lh_energy_strained}. The experimental measurements of the strain-dependent Ge direct band gap of Ref.~\onlinecite{ElKurdi2010} were performed at room temperature. In order to compare these data to our zero-temperature DFT calculations, we have applied a rigid energy shift to the experimental data so that the measured and calculated direct band gaps coincide at zero strain. We then note excellent quantitative agreement between theory and experiment, with our DFT-based deformation potential theory calculations quantitatively describing the tensile strain-induced reduction of the direct band gap.


For pristine Ge, our calculations predict that pseudomorphic tensile strain can produce an indirect L$_{6}^{+}$-$\Gamma_{8}^{+}$ to direct $\Gamma_{7}^{-}$-$\Gamma_{8}^{+}$ band gap transition -- highlighted by the crossing of the dashed green and red lines in Fig.~\ref{fig:Ge_and_Sn_strained}(a) -- at in-plane tensile strain $\epsilon_{xx} \approx 1.4$\%. This transition is driven by a combination of (i) a downward shift of the $\Gamma_{7}^{-}$ zone-centre CB edge energy due to the hydrostatic component of the pseudomorphic tensile strain (cf.~Eq.~\eqref{eq:cb_energy_strained}), and (ii) an upward shift in energy of the LH-like $\Gamma_{8}^{+}$ VB edge states due to the biaxial component of the pseudomorphic tensile strain (cf.~Eq.~\eqref{eq:lh_energy_strained}). We note that this is comparable to previous predictions of an indirect- to direct-gap transition for $\epsilon_{xx}$ in the range 1.5 -- 1.9\%. \cite{Gupta2013, Virgilio2013,ElKurdi2010a} For larger tensile strain, we observe further narrowing of the direct band gap, and predict that the band gap  can be closed in Ge for tensile strain $\epsilon_{xx} \approx 4.4$\%. Under compressive strain we compute an increase (decrease) of the direct $\Gamma_{7}^{-}$-$\Gamma_{8}^{+}$ (indirect L$_{6}^{+}$-$\Gamma_{8}^{+}$) band gap. Here, we emphasise that we are considering Ge under applied pseudomorphic strain. Under purely hydrostatic applied compression it would be expected that both the direct and direct band gaps increase in magnitude, as reflected by the negative values of the associated band gap hydrostatic deformation potentials (cf.~Table~\ref{tab:dft_benchmark}). Here, under compressive pseudomorphic strain, the upward energy shift of the HH-like $\Gamma_{8}^{+}$ VB edge states at fixed strain is larger than the upward energy shift of the L$_{6}^{+}$ CB edge states, leading to a net reduction of the indirect band gap.

Considering our calculated results for pristine $\alpha$-Sn in Fig.~\ref{fig:Ge_and_Sn_strained}(b), we note similar behaviour with good quantitative agreement between the TB-mBJ DFT- and deformation potential theory-calculated strain-dependent band gaps. We note, however, that while it would be expected that it should be possible to open the semimetallic (inverted) $\Gamma_{7}^{-}$-$\Gamma_{8}^{+}$ energy gap under applied hydrostatic pressure, the splitting of the HH- and LH-like $\Gamma_{8}^{+}$ states in the presence of symmetry-breaking pseudomorphic strain prevents this. Specifically, we compute that the biaxial component of the pseudomorphic strain pushes the HH-like $\Gamma_{8}^{+}$ states higher in energy at a rate which is approximately equal to the upward energy shift of the $\Gamma_{7}^{-}$ states due to the hydrostatic component of the strain. As such, the computed inverted direct band gap in Fig.~\ref{fig:Ge_and_Sn_strained}(b) -- closed green circles and solid green line -- remains approximately constant as a function of compressive in-plane strain. Therefore, our calculations suggest that while it is in principle possible to close the energy band gap in highly tensile strained pseudomorphic Ge, it is not possible to open a band gap for realistic strain values in bulk-like $\alpha$-Sn pseudomorphic strained layers.


\begin{figure*}
\centering
\includegraphics[width=0.90\textwidth]{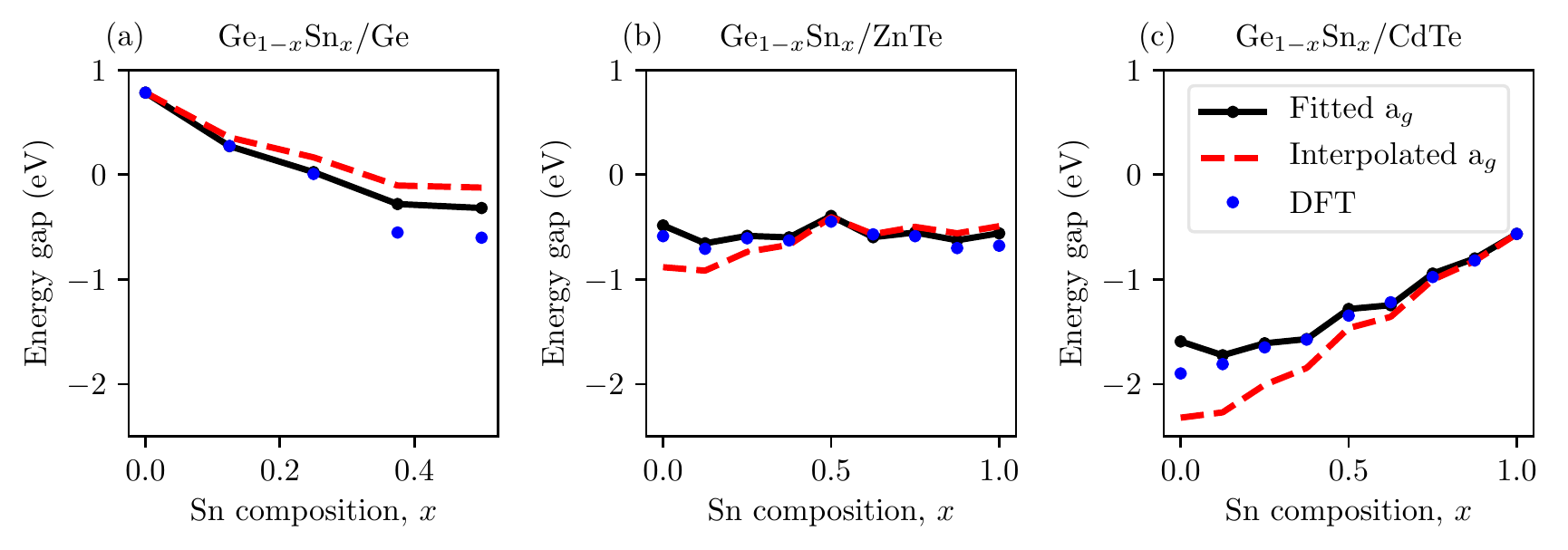}
\caption{Calculated $\Gamma_{7}^{-}$-$\Gamma_{8}^{+}$ band gap as a function of Sn composition $x$, for pseudomorphically strained Ge$_{1-x}$Sn$_{x}$ SQSs grown on [001]-oriented (a) Ge, (b) ZnTe, and (c) CdTe substrates. Closed blue circles denote direct meta-GGA DFT calculations. Dashed red and solid black lines show the band gaps calculated using Eqs.~\eqref{eq:cb_energy_strained} --~\eqref{eq:so_energy_strained}, respectively employing a linearly interpolated and best-fit value for $a_{g} ( \Gamma_{7}^{-}\text{-}\Gamma_{8}^{+} )$.
\label{fig:GeSn_strained_band_gap_vs_dpt}}
\end{figure*}


Next, we investigate the strain-dependent band gaps of pseudomorphic Ge$_{1-x}$Sn$_{x}$ alloys grown on Ge, ZnTe and CdTe, and analyse how alloying and strain can be combined to engineer semiconducting or semimetallic epitaxial layers. Closed blue circles in Figs.~\ref{fig:GeSn_strained_band_gap_vs_dpt}(a),~\ref{fig:GeSn_strained_band_gap_vs_dpt}(b) and~\ref{fig:GeSn_strained_band_gap_vs_dpt}(c) respectively show the relevant DFT-calculated $\Gamma_{7}^{-}$-$\Gamma_{8}^{+}$ energy gaps for pseudomorphically strained (a) Ge$_{1-x}$Sn$_{x}$/Ge, (b) Ge$_{1-x}$Sn$_{x}$/ZnTe and (c) Ge$_{1-x}$Sn$_{x}$/CdTe alloy SQSs. Results for Ge$_{1-x}$Sn$_{x}$/Ge are shown only for $x < 0.50$ as 64-atom SQSs having larger Sn compositions become amorphous when the ionic positions are allowed to relax. \cite{Miscibility}

These DFT calculations are compared to deformation potential theory calculations in which the hydrostatic band gap deformation potential $a_{g}$ is obtained via (i) linear interpolation between the values for Ge and $\alpha$-Sn at each value of $x$ (dashed red lines), or (ii) fitting to the results of the DFT calculations (solid black lines). In both cases, the VB edge axial deformation potential $b ( \Gamma_{8}^{+} )$ for the alloy is determined by linear interpolation between the corresponding values for Ge and $\alpha$-Sn (cf.~Table~\ref{tab:dft_benchmark}). The solid black lines in Figs.~\ref{fig:GeSn_strained_band_gap_vs_dpt}(a),~\ref{fig:GeSn_strained_band_gap_vs_dpt}(b) and~\ref{fig:GeSn_strained_band_gap_vs_dpt}(c) correspond to a best fit value $a_{g} ( \Gamma_{7}^{-}\text{-}\Gamma_{8}^{+} ) = -5.34$ eV, obtained by fitting to the DFT-calculated inverted direct energy gap of Ge$_{1-x}$Sn$_{x}$/ZnTe in Fig.~\ref{fig:GeSn_strained_band_gap_vs_dpt}(b). We note that this fitting is relatively insensitive to either Sn composition or choice of substrate, with similar values obtained by fitting to the DFT results for Ge$_{1-x}$Sn$_{x}$/Ge or Ge$_{1-x}$Sn$_{x}$/CdTe. We note that this composition-independent best-fit value for $a_{g} ( \Gamma_{7}^{-}\text{-}\Gamma_{8}^{+} )$ produces good quantitative agreement between the full DFT and model deformation potential calculations, tending only to break down in the presence of large in-plane strains $\vert \epsilon_{xx} \vert \gtrsim 5$\% -- i.e.~at high and low $x$ in Figs.~\ref{fig:GeSn_strained_band_gap_vs_dpt}(a) and~\ref{fig:GeSn_strained_band_gap_vs_dpt}(c) respectively -- where linear-in-strain deformation potential theory is expected to lose accuracy.


The best-fit value $a_{g} = -5.34$ eV for the hydrostatic deformation potential associated with the Ge$_{1-x}$Sn$_{x}$ direct band gap is not intermediate between the values of $-9.54$ and $-6.68$ eV calculated respectively for the $\Gamma_{7}^{-}$-$\Gamma_{8}^{+}$ band gaps of Ge and $\alpha$-Sn (cf.~Table~\ref{tab:dft_benchmark}). This suggests that alloy-related band mixing effects play a key role in determining the nature of the band gap in Ge$_{1-x}$Sn$_{x}$ alloys, as the best fit value lies $> 1$ eV outside of the range expected from a virtual crystal-type linear interpolation which neglects potential band hybridisation. Indeed, our best-fit value for $a_{g} ( \Gamma_{7}^{-}\text{-}\Gamma_{8}^{+} )$ is instead intermediate between the values associated with the direct $\Gamma_{7}^{-}$-$\Gamma_{8}^{+}$ and indirect L$_{6}^{+}$-$\Gamma_{8}^{+}$ band gaps of Ge, reflecting that band folding in our 64-atom supercell calculations allows for hybridisation of Ge $\Gamma$- and L-point states in response to Sn incorporation. This is in agreement with recent theoretical analysis \cite{OHalloran2019} suggesting that Sn incorporation in Ge drives strong hybridisation of Ge $\Gamma$- and L-point CB edge states, leading to a band gap that is neither purely indirect nor direct in nature, but which evolves continuously from having indirect to direct character via composition-dependent alloy band mixing. For $x \lesssim 0.10$ this conclusion is supported by pressure-dependent measurements, \cite{Eales2019} which demonstrate that the pressure coefficient $\frac{ d E_{g} }{ dP }$ associated with the Ge$_{1-x}$Sn$_{x}$ fundamental band gap is that of the indirect L$_{6}^{+}$-$\Gamma_{8}^{+}$ Ge band gap at $x = 0$ (4.3 meV kbar$^{-1}$), and increases continuously with increasing $x$ until it reaches a value close to that associated with the direct $\Gamma_{7}^{-}$-$\Gamma_{8}^{+}$ band gap of Ge by $x \approx 0.10$\ (12.9 meV kbar$^{-1}$). Converting our best-fit value $a_{g} = -5.34$ eV to a pressure coefficient (cf.~Sec.~\ref{sec:theory_strain}) -- via computation of the alloy bulk modulus based on interpolation of the elastic constants using the bowing parameters of Sec.~\ref{sec:theory_dft} -- we indeed obtain an intermediate value $\frac{ d E_{g} }{ dP } \approx 7$ meV kbar$^{-1}$. We note that the limited band folding present in the $\lesssim 10^{2}$-atom supercells accessible using meta-GGA DFT calculations limits the number of states that can hybridise in response to alloying. As such, the observed Sn-induced $\Gamma$-L band mixing observed in our calculations persists across the entire composition range as an artefact of the supercells employed in our calculations, whereas experimental measurements suggest that alloy band mixing effects are most pronounced for $x \lesssim 0.10$. Indeed, we have recently described the presence of similar spurious hybridisation effects in alloy supercell calculations for Ge$_{1-x}$Pb$_{x}$ alloys. \cite{Broderick_arXiv_2019} Nonetheless, our calculations then support the emerging re-evaluation of the nature of the indirect- to direct-gap transition in Ge$_{1-x}$Sn$_{x}$ alloys in terms of alloy band mixing effects, \cite{OHalloran2019,Eales2019} which have largely been neglected in previous analyses.

\begin{figure}
\centering
\includegraphics[width=0.90\columnwidth]{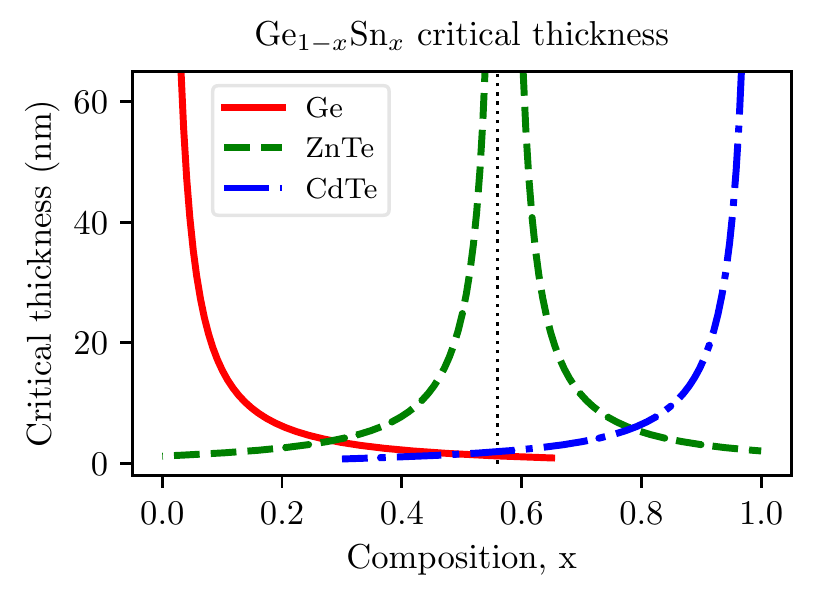}
\caption{Calculated Sn composition-dependent critical thickness $t_{c}$ of pseudomorphically strained Ge$_{1-x}$Sn$_{x}$ grown on [001]-oriented Ge (solid red line), ZnTe (dashed green line), or CdTe (dash-dotted blue line) substrates.}
\label{fig:critical_thickness}
\end{figure}

To compute the critical thickness $t_{c}$ for Ge$_{1-x}$Sn$_{x}$ grown on Ge, ZnTe and CdTe substrates we use Eq.~\eqref{eq:critical_thickness} with the DFT-calculated lattice and elastic constants listed in Table~\ref{tab:dft_benchmark}. The results of these calculations are shown in Fig.~\ref{fig:critical_thickness} using solid red, dashed green and dash-dotted blue lines, respectively. Firstly, we calculate critical thicknesses at $\vert \epsilon_{xx} \vert = 1$\% in order to provide guideline strain-thickness limit estimations. Our estimated strain-thickness limits are, respectively, $t_{c} \times \vert \epsilon_{xx} \vert = 22.8$ nm \% and 24.4 nm \% for Ge$_{1-x}$Sn$_{x}$ grown on Ge and CdTe. For Ge$_{1-x}$Sn$_{x}$/ZnTe, which is under compressive (tensile) in-plane strain for $x \geq 54$\% ($x \leq 0.54$), we estimate a strain-thickness limit $t_{c} \times \vert \epsilon_{xx} \vert = 23.3$ nm \% (23.5 nm \%). To compare the different substrates we choose a reference critical thickness of 5 nm, as a representative thickness along [001] of a thin film in which quantum confinement effects can open a band gap in semimetallic Ge$_{1-x}$Sn$_{x}$. For Ge$_{1-x}$Sn$_{x}$/Ge we compute that the critical thickness reduces to 5 nm for Sn composition $x = 0.234$, corresponding to an in-plane compressive strain of magnitude $\vert \epsilon_{xx} \vert = 3.4$\%. In previous analysis, \cite{Miscibility} we predicted based on LDA-calculated SQS total energies that it is energetically favourable for Ge$_{1-x}$Sn$_{x}$ to become amorphous for $x \gtrsim 0.56$. The calculations of Ref.~\onlinecite{Miscibility} were performed for the same 64-atom SQSs considered in this work: these $2 \times 2 \times 2$ simple cubic supercells have a thickness of $2 \times a(x) = 1.2$ nm, where $a (x)$ is the relaxed alloy lattice constant along [001]. Indeed, we calculate $t_{c} \leq 2 \, a(x)$ for $x \geq 0.566$. This excellent quantitative agreement between DFT alloy SQS total energy calculations and the critical thickness computed via Eq.~\eqref{eq:critical_thickness} provides confidence in our predicted strain-thickness limits for Ge$_{1-x}$Sn$_{x}$. For Ge$_{1-x}$Sn$_{x}$/ZnTe we compute that $t_{c}$ is reduced to 5 nm for $x = 0.334$ ($x = 0.808$), corresponding to a tensile (compressive) in-plane strain of magnitude $\vert \epsilon_{xx} \vert = 3.1$\% ($\vert \epsilon_{xx} \vert = 3.4$\%). Finally, for tensile strained Ge$_{1-x}$Sn$_{x}$/CdTe we compute $t_{c} = 5$ nm for $x = 0.759$, corresponding to a tensile in-plane strain of magnitude $\vert \epsilon_{xx} \vert = 2.8$\%. We therefore note that growing Ge$_{1-x}$Sn$_{x}$ on these three substrates allows $t_{c} \geq 5$ nm to be achieved across almost the entire composition range, with the exclusion of a gap for $0.234 \leq x \leq 0.334$, corresponding to relaxed lattice constants in the range $a (x) \approx 5.84$ -- 5.92 \AA.


We are now in a position to consider quantitatively the implications of the electronic structure and critical thickness for engineering the semimetal-to-semiconductor transition for the three substrates considered in this work. Growth of Ge$_{1-x}$Sn$_{x}$ on CdTe substrates increases the magnitude of the inverted $\Gamma_{7}^{-}$-$\Gamma_{8}^{+}$ energy gap with decreasing $x$ (cf.~Fig.~\ref{fig:GeSn_band_gap_strained}). This, combined with the associated reduction in critical thickness (cf.~Fig.~\ref{fig:critical_thickness}), suggests that growth of Ge$_{1-x}$Sn$_{x}$ on substrates having lattice constant close to that of $\alpha$-Sn is not a suitable approach to produce semimetallic thin films in which a band gap could be opened below a given film thickness. Pseudomorphic growth of Ge$_{1-x}$Sn$_{x}$ on Ge requires Sn composition $x \gtrsim 0.26$ to close the alloy band gap and achieve a semimetallic alloy, compared to $x \approx 0.21$ in a relaxed alloy. In this composition range, we compute Ge$_{1-x}$Sn$_{x}$/Ge critical thickness $t_{c} \lesssim 5$ nm, suggesting that it is challenging to grow sufficiently thick defect-free films in order to achieve semimetallic materials, and hence the film would be semiconducting for the film thicknesses that could be achieved, preventing band gap modulation via variation of the film thickness. Finally, growth of Ge$_{1-x}$Sn$_{x}$ on a ZnTe substrate having a lattice constant intermediate between Ge and $\alpha$-Sn produces an inverted-gap semimetallic alloy, with the magnitude of the inverted energy gap being relatively insensitive to Sn composition $x$. This stability of the inverted band gap with Sn composition $x$ for Ge$_{1-x}$Sn$_{x}$/ZnTe could allow pseudomorphic growth of alloys having Sn composition close to lattice-matching to a ZnTe substrate for $x = 0.54$. As such, growth of Ge$_{1-x}$Sn$_{x}$ on ZnTe provides a route to circumvent critical thickness limitations simultaneously allowing, at fixed Sn composition, for growth of films having thicker semimetallic and thinner semiconducting regions. Our calculations suggest that Ge$_{1-x}$Sn$_{x}$/ZnTe layers have $t_{c} > 10$ nm for $0.42 \lesssim x \lesssim 0.69$, with the lower end of this composition range being more favourable for epitaxial growth in order to mitigate deleterious Sn segregation arising due to reduced alloy miscibility at higher Sn compositions. \cite{Miscibility,Suzuki2016,Tsukamoto2015}


\begin{figure}
\centering
\includegraphics[width=0.90\columnwidth]{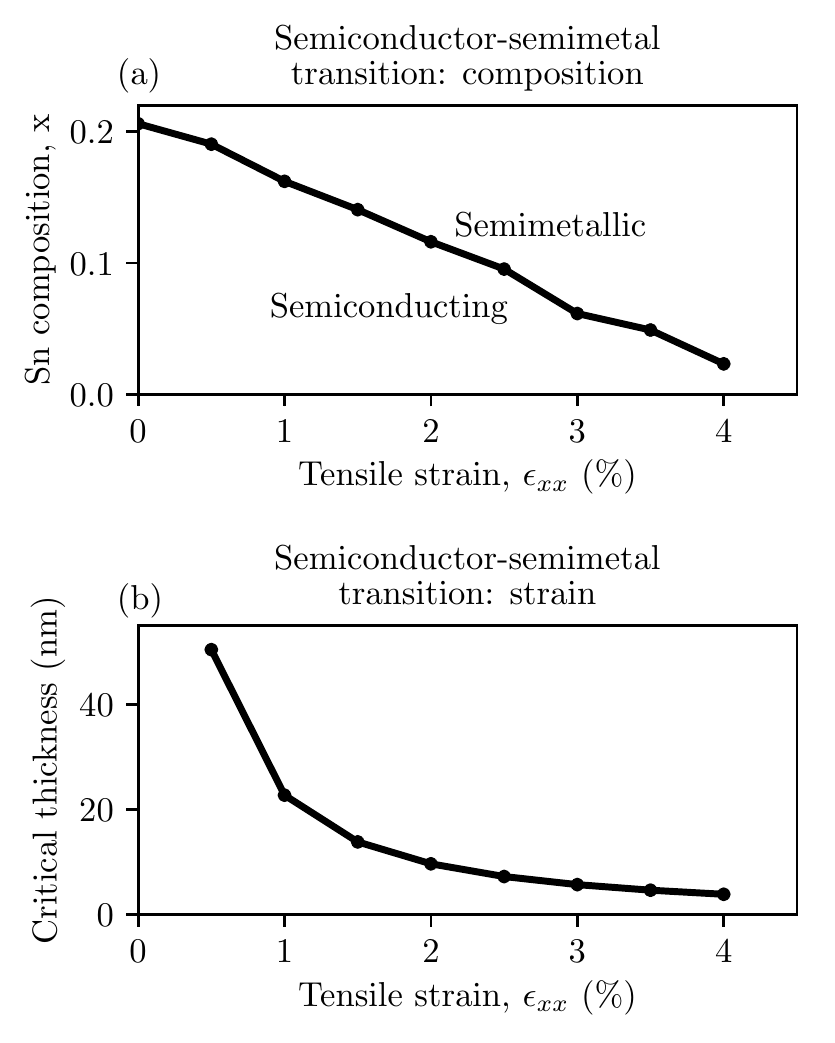}
\caption{(a) Sn composition $x$ at which the band gap of [001]-oriented pseudomorphic Ge$_{1-x}$Sn$_{x}$ closes -- i.e.~the alloy becomes semimetallic -- as a function of in-plane tensile strain $\epsilon_{xx}$, demonstrating that the Sn composition required to close the alloy band gap can be strongly reduced via choice of substrate. (b) Critical thickness associated with the alloy Sn compositions of (a).}
\label{fig:semimetallic_transition}
\end{figure}


Growth of Ge$_{1-x}$Sn$_{x}$/ZnTe having Sn compositions $x \gtrsim 0.42$\ is demanding from the perspective of epitaxial growth. As such, it is also of interest to quantify the degree to which tensile strain can be exploited to lower the minimum Sn composition at which the alloy becomes semimetallic. This is similar in spirit to recent work on low $x$ prototype Ge$_{1-x}$Sn$_{x}$ lasers, where Sn incorporation and tensile strain are simultaneously employed to drive the emergence of a direct band gap. \cite{Rainko_NP_2020} Reducing Sn composition reduces Sn segregation during growth, and promotes enhanced crystalline quality by enabling higher temperature growth. However, reduction in Sn composition at fixed strain will increase the magnitude of the band gap in the semiconducting regime. Application of tensile strain to reduce the band gap also reduces the critical thickness, leading to a trade-off for epitaxial growth of semimetallic Ge$_{1-x}$Sn$_{x}$. In practice, simultaneous introduction of tensile strain and a reduction of Sn composition required to close the band gap can be achieved via growth on substrates having lattice constant intermediate between those associated with the relaxed Ge$_{1-x}$Sn$_{x}$ alloy and ZnTe.

We have calculated the minimum Sn composition at which the alloy becomes semimetallic as a function of applied tensile pseudomorphic strain for the same SQSs considered above. The results of these calculations are summarised in Figs.~\ref{fig:semimetallic_transition}(a) and~\ref{fig:semimetallic_transition}(b) which show, as a function of in-plane tensile strain, (a) the Sn composition $x$ at which the alloy band gap closes, and (b) the critical thickness $t_{c}$ corresponding to the Sn composition and tensile strain in (a). Figure~\ref{fig:semimetallic_transition}(a) therefore summarises as a function of Sn composition and tensile strain the conditions under which pseudomorphic Ge$_{1-x}$Sn$_{x}$ alloys are predicted to be semiconducting or semimetallic. At zero strain, corresponding to the relaxed alloy calculations of Fig.~\ref{fig:GeSn_band_gap_unstrained}, we recall that $x \approx 0.21$ is required to close the alloy band gap. The Sn composition at which the band gap closes is calculated to decrease approximately linearly with applied tensile strain, by $\approx 4.6$\% Sn per \% tensile strain (i.e.~in-plane lattice mismatch $\epsilon_{xx}$). We note that extrapolation to $x = 0$ in Fig.~\ref{fig:semimetallic_transition}(a) gives $\epsilon_{xx} \approx 4.4$\%, corresponding to the closing of the direct band gap of Ge under pseudomorphic tensile strain (cf.~Fig.~\ref{fig:Ge_and_Sn_strained}(a)). Examining Fig.~\ref{fig:semimetallic_transition}(b) we find $t_{c} \gtrsim 10$ nm for $\epsilon_{xx} \lesssim 2$\%, corresponding in Fig.~\ref{fig:semimetallic_transition}(a) to Sn compositions $x \gtrsim 0.11$. As such, we predict that the minimum Sn composition required to close the alloy band gap can be reduced by a factor of approximately two in a tensile strained pseudomorphic Ge$_{1-x}$Sn$_{x}$ layer compared to a relaxed alloy, assuming that the layer is sufficiently thick to prevent opening of a band gap via quantum confinement.

The strained lattice constants of Figs.~\ref{fig:semimetallic_transition}(a) and~\ref{fig:semimetallic_transition}(b) lie in the range 5.80 -- 5.89~\AA, lower than the lattice constant of ZnTe. This suggests additional paths to engineer the minimum Sn composition at which the alloy becomes semimetallic: pseudomorphic growth of Ge$_{1-x}$Sn$_{x}$ on InP substrates, or metamorphic growth on relaxed In$_{y}$Ga$_{1-y}$As metamorphic buffer layers. \cite{Lin2012} InP has a lattice constant of $5.87$~\AA, \cite{Vurgaftman2001} lying within the range described above. In$_{y}$Ga$_{1-y}$As has a lattice constant varying between 5.65~\AA~($y = 0$) and 6.06~\AA~($y = 1$), \cite{Haynes2014} which covers the entire range described above, allowing for lattice matching to relaxed Ge$_{1-x}$Sn$_{x}$ alloys up to $x \approx 0.50$. Finally, Figs.~\ref{fig:GeSn_band_gap_strained} and~\ref{fig:critical_thickness} also suggest an alternative strategy for producing Ge$_{1-x}$Sn$_{x}$ thin films to engineer a semimetal-to-semiconductor transition by varying alloy film thickness. Pseudomorphic growth of a sufficiently thick, low $x$ Ge$_{1-x}$Sn$_{x}$/Ge metamorphic buffer layer would exhibit semiconducting behaviour. Grown sufficiently thick to allow for relaxation, such a layer will possess an increased lattice constant relative to the Ge substrate. Subsequent growth on top of this buffer layer with increased $x$ will increase the critical thickness relative to Ge$_{1-x}$Sn$_{x}$/Ge, pushing a given critical thickness to higher $x$. In this manner, thicker semimetallic Ge$_{1-x}$Sn$_{x}$ films grown on a lower $x$ metamorphic buffer may be achievable. Alternatively, thinner unrelaxed buffer layers with increased $x$ would also be suitable, as compressive strain and quantum confinement contribute to opening a band gap, allowing for a semiconducting virtual substrate. With a reduced band gap relative to the Ge substrate and buffer layer, electrical measurements can indicate if a semimetal-to-semiconductor transition can be achieved using such a growth scheme.


\section{Conclusions}
\label{sec:conclusions}


In summary, we have presented a theoretical analysis of the electronic structure of Ge$_{1-x}$Sn$_{x}$ across the full alloy composition range. Using DFT calculations for SQS alloy supercells we have quantified the nature and evolution of the band gap of free-standing and pseudomorphically strained Ge$_{1-x}$Sn$_{x}$ with Sn composition $x$, and interpreted the electronic structure via (i) direct quantitative analysis of key alloy eigenstates, and (ii) model analysis based on deformation potential theory. Based on parameters extracted from DFT calculations, we additionally employed continuum elasticity theory to provide estimates of critical thickness limits associated with pseudomorphic growth of Ge$_{1-x}$Sn$_{x}$ epitaxial layers on Ge, ZnTe and CdTe substrates (with ZnTe and CdTe chosen for having lattice constant respectively close to that of zb-GeSn and $\alpha$-Sn). The results of our electronic structure calculations were found to be in good quantitative agreement with experimental measurements across the range of Sn compositions for which data are currently available. Furthermore, our predicted critical thickness limits for pseudomorphically strained Ge$_{1-x}$Sn$_{x}$ alloys were found to be in quantitative agreement with the results of our first principles analysis of thermodynamic stability in Ge$_{1-x}$Sn$_{x}$, \cite{Miscibility} predicting that it is energetically favourable for epitaxial Ge$_{1-x}$Sn$_{x}$/Ge layers to become amorphous for $x \gtrsim 0.56$.


In the low Sn composition range $x \lesssim 0.10$, our results firstly confirmed the presence of an indirect- to direct-gap transition, as described widely in the literature. The calculated values of the hydrostatic deformation potential (pressure coefficient) associated with the fundamental alloy band gap in this composition range were found to be intermediate between those associated with the indirect fundamental L$_{6}^{+}$-$\Gamma_{8}^{+}$, and direct $\Gamma_{7}^{-}$-$\Gamma_{8}^{+}$ band gaps of Ge. This supports emerging evidence suggesting that Sn-induced hybridisation of Ge CB states plays a key role in characterising the indirect to direct-gap transition, which will have significant consequences for properties relevant to proposed device applications at low $x$. For $x \gtrsim 0.10$ we demonstrated that the minimum Sn composition required to close the alloy band gap can be engineered via both alloying and pseudomorphic strain. We predict that free-standing Ge$_{1-x}$Sn$_{x}$ becomes semimetallic at $x \approx 0.21$, and that combining alloying of Ge and Sn with strain engineering allows to significantly reduce the magnitude of the inverted (negative) $\Gamma_{7}^{-}$-$\Gamma_{8}^{+}$ energy that must be overcome by quantum confinement to open a band gap in semimetallic Ge$_{1-x}$Sn$_{x}$. For tensile strained pseudomorphic Ge$_{1-x}$Sn$_{x}$, we estimate that the Sn composition required to reduce the inverted $\Gamma_{7}^{-}$-$\Gamma_{8}^{+}$ energy separation can be reduced to $x \approx 0.11$ in Ge$_{1-x}$Sn$_{x}$ pseudomorphic layers having critical thickness $t_{c} \approx 10$ nm.


The realisation of recently proposed nanoscale intrinsic, confinement-modulated-gap diodes and transistors relies on the availability of materials for which the semimetal-to-semiconductor transition can be engineered to create thicker semimetallic regions adjacent to thinner semiconducting regions. In this respect, Ge$_{1-x}$Sn$_{x}$ alloys offer opportunities to engineer the electronic structure to meet novel device requirements. One advantage is the reduction of the magnitude of the inverted semimetallic energy gap to be overcome by quantum confinement in order to induce a band gap in semimetallic Ge$_{1-x}$Sn$_{x}$, compared to that in $\alpha$-Sn. Reducing this inverted energy gap relaxes constraints on the minimum film thickness at which the semimetal-to-semiconductor transition can be observed. The possibility to explore pseudomorphic growth on different substrates and metamorphic buffer layers provides opportunities to combine alloy composition and strain engineering to control the Sn composition at which Ge$_{1-x}$Sn$_{x}$ becomes semimetallic. We note that further possibilities for band gap engineering in thin films and nanowires are also offered by, e.g., choice of surface termination, making Ge$_{1-x}$Sn$_{x}$ nanostructures a promising platform for electronics applications. Overall, we conclude that combining pseudomorphic strain with alloying in Ge$_{1-x}$Sn$_{x}$ presents rich opportunities for band structure engineering, presenting opportunities for novel applications in electronic and photonic devices at higher Sn compositions.


\begin{acknowledgments}

This work was supported by Science Foundation Ireland (SFI; project nos.~13/IA/1956 and 15/IA/3082), by the National University of Ireland (NUI; via the Post-Doctoral Fellowship in the Sciences, held by C.A.B.), with partial support from the Nottingham-Ningbo Materials Institute and National Natural Science Foundation of China with a project code 61974079.

\end{acknowledgments}



\begin{thebibliography}{91}
\expandafter\ifx\csname natexlab\endcsname\relax\def\natexlab#1{#1}\fi
\expandafter\ifx\csname bibnamefont\endcsname\relax
  \def\bibnamefont#1{#1}\fi
\expandafter\ifx\csname bibfnamefont\endcsname\relax
  \def\bibfnamefont#1{#1}\fi
\expandafter\ifx\csname citenamefont\endcsname\relax
  \def\citenamefont#1{#1}\fi
\expandafter\ifx\csname url\endcsname\relax
  \def\url#1{\texttt{#1}}\fi
\expandafter\ifx\csname urlprefix\endcsname\relax\def\urlprefix{URL }\fi
\providecommand{\bibinfo}[2]{#2}
\providecommand{\eprint}[2][]{\url{#2}}

\bibitem[{\citenamefont{Darmody et~al.}(2015)\citenamefont{Darmody, Ettisserry,
  Goldsman, and Dhar}}]{DarmodyEttisserryGoldsmanEtAl2015}
\bibinfo{author}{\bibfnamefont{C.}~\bibnamefont{Darmody}},
  \bibinfo{author}{\bibfnamefont{D.~P.} \bibnamefont{Ettisserry}},
  \bibinfo{author}{\bibfnamefont{N.}~\bibnamefont{Goldsman}}, \bibnamefont{and}
  \bibinfo{author}{\bibfnamefont{N.~K.} \bibnamefont{Dhar}}, in
  \emph{\bibinfo{booktitle}{International Conference on Simulation of
  Semiconductor Processes and Devices ({SISPAD})}} (\bibinfo{year}{2015}).

\bibitem[{\citenamefont{Sau and Cohen}(2007)}]{SauCohen2007}
\bibinfo{author}{\bibfnamefont{J.~D.} \bibnamefont{Sau}} \bibnamefont{and}
  \bibinfo{author}{\bibfnamefont{M.~L.} \bibnamefont{Cohen}},
  \bibinfo{journal}{Phys.~Rev.~B} \textbf{\bibinfo{volume}{75}}
  (\bibinfo{year}{2007}).

\bibitem[{\citenamefont{Dutt et~al.}(2013)\citenamefont{Dutt, Lin, Sukhdeo,
  Vulovi\'{c}, Gupta, Nam, Saraswat, and Harris}}]{DuttLinSukhdeoEtAl2013}
\bibinfo{author}{\bibfnamefont{B.}~\bibnamefont{Dutt}},
  \bibinfo{author}{\bibfnamefont{H.}~\bibnamefont{Lin}},
  \bibinfo{author}{\bibfnamefont{D.~S.} \bibnamefont{Sukhdeo}},
  \bibinfo{author}{\bibfnamefont{B.~M.} \bibnamefont{Vulovi\'{c}}},
  \bibinfo{author}{\bibfnamefont{S.}~\bibnamefont{Gupta}},
  \bibinfo{author}{\bibfnamefont{D.}~\bibnamefont{Nam}},
  \bibinfo{author}{\bibfnamefont{K.~C.} \bibnamefont{Saraswat}},
  \bibnamefont{and} \bibinfo{author}{\bibfnamefont{J.~S.}
  \bibnamefont{Harris}}, \bibinfo{journal}{IEEE J.~Sel.~Top.~Quantum Electron.}
  \textbf{\bibinfo{volume}{19}}, \bibinfo{pages}{1502706}
  (\bibinfo{year}{2013}).

\bibitem[{\citenamefont{Low et~al.}(2012)\citenamefont{Low, Yang, Han, Fan, and
  Yeo}}]{Low2012}
\bibinfo{author}{\bibfnamefont{K.~L.} \bibnamefont{Low}},
  \bibinfo{author}{\bibfnamefont{Y.}~\bibnamefont{Yang}},
  \bibinfo{author}{\bibfnamefont{G.}~\bibnamefont{Han}},
  \bibinfo{author}{\bibfnamefont{W.}~\bibnamefont{Fan}}, \bibnamefont{and}
  \bibinfo{author}{\bibfnamefont{Y.-C.} \bibnamefont{Yeo}},
  \bibinfo{journal}{J.~Appl.~Phys.} \textbf{\bibinfo{volume}{112}},
  \bibinfo{pages}{103715} (\bibinfo{year}{2012}).

\bibitem[{\citenamefont{Doherty et~al.}(2020)\citenamefont{Doherty, Biswas,
  Galluccio, Broderick, Garcia-Gil, Duffy, O'Reilly, and
  Holmes}}]{Doherty_CM_2020}
\bibinfo{author}{\bibfnamefont{J.}~\bibnamefont{Doherty}},
  \bibinfo{author}{\bibfnamefont{S.}~\bibnamefont{Biswas}},
  \bibinfo{author}{\bibfnamefont{E.}~\bibnamefont{Galluccio}},
  \bibinfo{author}{\bibfnamefont{C.~A.} \bibnamefont{Broderick}},
  \bibinfo{author}{\bibfnamefont{A.}~\bibnamefont{Garcia-Gil}},
  \bibinfo{author}{\bibfnamefont{R.}~\bibnamefont{Duffy}},
  \bibinfo{author}{\bibfnamefont{E.~P.} \bibnamefont{O'Reilly}},
  \bibnamefont{and} \bibinfo{author}{\bibfnamefont{J.~D.}
  \bibnamefont{Holmes}}, \bibinfo{journal}{Chem.~Mater.}
  \textbf{\bibinfo{volume}{32}}, \bibinfo{pages}{4383} (\bibinfo{year}{2020}).

\bibitem[{\citenamefont{Homewood and Louren\c{c}o}(2015)}]{Homewood2015}
\bibinfo{author}{\bibfnamefont{K.~P.} \bibnamefont{Homewood}} \bibnamefont{and}
  \bibinfo{author}{\bibfnamefont{M.~A.} \bibnamefont{Louren\c{c}o}},
  \bibinfo{journal}{Nature Photonics} \textbf{\bibinfo{volume}{9}},
  \bibinfo{pages}{78} (\bibinfo{year}{2015}).

\bibitem[{\citenamefont{Wirths et~al.}(2015)\citenamefont{Wirths, Geiger,
  von~den Driesch, Mussler, Stoica, Mantl, Ikonic, Luysberg, Chiussi, Hartmann
  et~al.}}]{Wirths2015}
\bibinfo{author}{\bibfnamefont{S.}~\bibnamefont{Wirths}},
  \bibinfo{author}{\bibfnamefont{R.}~\bibnamefont{Geiger}},
  \bibinfo{author}{\bibfnamefont{N.}~\bibnamefont{von~den Driesch}},
  \bibinfo{author}{\bibfnamefont{G.}~\bibnamefont{Mussler}},
  \bibinfo{author}{\bibfnamefont{T.}~\bibnamefont{Stoica}},
  \bibinfo{author}{\bibfnamefont{S.}~\bibnamefont{Mantl}},
  \bibinfo{author}{\bibfnamefont{Z.}~\bibnamefont{Ikonic}},
  \bibinfo{author}{\bibfnamefont{M.}~\bibnamefont{Luysberg}},
  \bibinfo{author}{\bibfnamefont{S.}~\bibnamefont{Chiussi}},
  \bibinfo{author}{\bibfnamefont{J.~M.} \bibnamefont{Hartmann}},
  \bibnamefont{et~al.}, \bibinfo{journal}{Nature Photonics}
  \textbf{\bibinfo{volume}{9}}, \bibinfo{pages}{88} (\bibinfo{year}{2015}).

\bibitem[{\citenamefont{Kasper and Oehme}(2015)}]{Kasper_2015}
\bibinfo{author}{\bibfnamefont{E.}~\bibnamefont{Kasper}} \bibnamefont{and}
  \bibinfo{author}{\bibfnamefont{M.}~\bibnamefont{Oehme}},
  \bibinfo{journal}{Jpn.~J.~Appl.~Phys.} \textbf{\bibinfo{volume}{54}},
  \bibinfo{pages}{04DG11} (\bibinfo{year}{2015}).

\bibitem[{\citenamefont{Yang et~al.}(2012)\citenamefont{Yang, Su, Guo, Wang,
  Gong, Wang, Low, Zhang, Xue, Cheng et~al.}}]{Yang_IEEEIEDM_2012}
\bibinfo{author}{\bibfnamefont{Y.}~\bibnamefont{Yang}},
  \bibinfo{author}{\bibfnamefont{S.}~\bibnamefont{Su}},
  \bibinfo{author}{\bibfnamefont{P.}~\bibnamefont{Guo}},
  \bibinfo{author}{\bibfnamefont{W.}~\bibnamefont{Wang}},
  \bibinfo{author}{\bibfnamefont{X.}~\bibnamefont{Gong}},
  \bibinfo{author}{\bibfnamefont{L.}~\bibnamefont{Wang}},
  \bibinfo{author}{\bibfnamefont{K.~L.} \bibnamefont{Low}},
  \bibinfo{author}{\bibfnamefont{G.}~\bibnamefont{Zhang}},
  \bibinfo{author}{\bibfnamefont{C.}~\bibnamefont{Xue}},
  \bibinfo{author}{\bibfnamefont{B.}~\bibnamefont{Cheng}},
  \bibnamefont{et~al.}, in \emph{\bibinfo{booktitle}{IEEE International
  Electron Devices Meeting (IEEE-IEDM)}} (\bibinfo{year}{2012}).

\bibitem[{\citenamefont{Liu et~al.}(2016)\citenamefont{Liu, Liang, Wang, Xiao,
  and Xu}}]{Liu2016SimulationOG}
\bibinfo{author}{\bibfnamefont{L.}~\bibnamefont{Liu}},
  \bibinfo{author}{\bibfnamefont{R.}~\bibnamefont{Liang}},
  \bibinfo{author}{\bibfnamefont{J.}~\bibnamefont{Wang}},
  \bibinfo{author}{\bibfnamefont{L.}~\bibnamefont{Xiao}}, \bibnamefont{and}
  \bibinfo{author}{\bibfnamefont{J.}~\bibnamefont{Xu}},
  \bibinfo{journal}{Appl.~Phys.~Express} \textbf{\bibinfo{volume}{9}},
  \bibinfo{pages}{091301} (\bibinfo{year}{2016}).

\bibitem[{\citenamefont{Sant and Schenk}(2015)}]{Sant_IEEEJEDS_2015}
\bibinfo{author}{\bibfnamefont{S.}~\bibnamefont{Sant}} \bibnamefont{and}
  \bibinfo{author}{\bibfnamefont{A.}~\bibnamefont{Schenk}},
  \bibinfo{journal}{IEEE J.~Electron Devices Soc.}
  \textbf{\bibinfo{volume}{3}}, \bibinfo{pages}{164 } (\bibinfo{year}{2015}).

\bibitem[{\citenamefont{Dunne et~al.}(2020)\citenamefont{Dunne, Broderick,
  Luisier, and O'Reilly}}]{Dunne2020}
\bibinfo{author}{\bibfnamefont{M.~D.} \bibnamefont{Dunne}},
  \bibinfo{author}{\bibfnamefont{C.~A.} \bibnamefont{Broderick}},
  \bibinfo{author}{\bibfnamefont{M.}~\bibnamefont{Luisier}}, \bibnamefont{and}
  \bibinfo{author}{\bibfnamefont{E.~P.} \bibnamefont{O'Reilly}}, in
  \emph{\bibinfo{booktitle}{International Conference on Numerical Simulation of
  Optoelectronic Devices (NUSOD)}} (\bibinfo{year}{2020}).

\bibitem[{\citenamefont{Moontragoon et~al.}(2007)\citenamefont{Moontragoon,
  Ikoni{\'{c}}, and Harrison}}]{MoontragoonIkonicHarrison2007}
\bibinfo{author}{\bibfnamefont{P.}~\bibnamefont{Moontragoon}},
  \bibinfo{author}{\bibfnamefont{Z.}~\bibnamefont{Ikoni{\'{c}}}},
  \bibnamefont{and} \bibinfo{author}{\bibfnamefont{P.}~\bibnamefont{Harrison}},
  \bibinfo{journal}{Semicond.~Sci.~Technol.} \textbf{\bibinfo{volume}{22}},
  \bibinfo{pages}{742} (\bibinfo{year}{2007}).

\bibitem[{\citenamefont{Jenkins and Dow}(1987)}]{JenkinsDow1987}
\bibinfo{author}{\bibfnamefont{D.~W.} \bibnamefont{Jenkins}} \bibnamefont{and}
  \bibinfo{author}{\bibfnamefont{J.~D.} \bibnamefont{Dow}},
  \bibinfo{journal}{Phys.~Rev.~B} \textbf{\bibinfo{volume}{36}},
  \bibinfo{pages}{7994} (\bibinfo{year}{1987}).

\bibitem[{\citenamefont{Polak et~al.}(2017)\citenamefont{Polak, Scharoch, and
  Kudrawiec}}]{Polak2017}
\bibinfo{author}{\bibfnamefont{M.~P.} \bibnamefont{Polak}},
  \bibinfo{author}{\bibfnamefont{P.}~\bibnamefont{Scharoch}}, \bibnamefont{and}
  \bibinfo{author}{\bibfnamefont{R.}~\bibnamefont{Kudrawiec}},
  \bibinfo{journal}{journal of Physics D: Applied Physics}
  \textbf{\bibinfo{volume}{50}}, \bibinfo{pages}{195103}
  (\bibinfo{year}{2017}).

\bibitem[{\citenamefont{Ansari et~al.}(2012)\citenamefont{Ansari, Fagas,
  Colinge, and Greer}}]{Ansari2012}
\bibinfo{author}{\bibfnamefont{L.}~\bibnamefont{Ansari}},
  \bibinfo{author}{\bibfnamefont{G.}~\bibnamefont{Fagas}},
  \bibinfo{author}{\bibfnamefont{J.-P.} \bibnamefont{Colinge}},
  \bibnamefont{and} \bibinfo{author}{\bibfnamefont{J.~C.} \bibnamefont{Greer}},
  \bibinfo{journal}{Nano Lett.} \textbf{\bibinfo{volume}{12}},
  \bibinfo{pages}{2222} (\bibinfo{year}{2012}).

\bibitem[{\citenamefont{Colinge and Greer}(2016)}]{nwt2017}
\bibinfo{author}{\bibfnamefont{J.~P.} \bibnamefont{Colinge}} \bibnamefont{and}
  \bibinfo{author}{\bibfnamefont{J.~C.} \bibnamefont{Greer}},
  \emph{\bibinfo{title}{Nanowire Transistors: Physics of Devices and Materials
  in One Dimension}} (\bibinfo{publisher}{Cambridge University Press},
  \bibinfo{address}{Cambridge}, \bibinfo{year}{2016}), ISBN
  \bibinfo{isbn}{9781107280779}.

\bibitem[{\citenamefont{Da et~al.}(2012)\citenamefont{Da, Lam, Samudra, Chin,
  and Liang}}]{DLS12}
\bibinfo{author}{\bibfnamefont{H.}~\bibnamefont{Da}},
  \bibinfo{author}{\bibfnamefont{K.}~\bibnamefont{Lam}},
  \bibinfo{author}{\bibfnamefont{G.}~\bibnamefont{Samudra}},
  \bibinfo{author}{\bibfnamefont{S.}~\bibnamefont{Chin}}, \bibnamefont{and}
  \bibinfo{author}{\bibfnamefont{G.}~\bibnamefont{Liang}},
  \bibinfo{journal}{IEEE Tran.~Electron.~Dev.} \textbf{\bibinfo{volume}{59}},
  \bibinfo{pages}{1454} (\bibinfo{year}{2012}).

\bibitem[{\citenamefont{Sanchez-Soares and Greer}(2016)}]{Sanchez-Soares2016}
\bibinfo{author}{\bibfnamefont{A.}~\bibnamefont{Sanchez-Soares}}
  \bibnamefont{and} \bibinfo{author}{\bibfnamefont{J.~C.} \bibnamefont{Greer}},
  \bibinfo{journal}{Nano Lett.} \textbf{\bibinfo{volume}{16}},
  \bibinfo{pages}{7639} (\bibinfo{year}{2016}).

\bibitem[{\citenamefont{Gity et~al.}(2017)\citenamefont{Gity, Ansari, Lanius,
  Sch\"{u}ffelgen, Mussler, Gr\"{u}tzmacher, and Greer}}]{GAL17}
\bibinfo{author}{\bibfnamefont{F.}~\bibnamefont{Gity}},
  \bibinfo{author}{\bibfnamefont{L.}~\bibnamefont{Ansari}},
  \bibinfo{author}{\bibfnamefont{M.}~\bibnamefont{Lanius}},
  \bibinfo{author}{\bibfnamefont{P.}~\bibnamefont{Sch\"{u}ffelgen}},
  \bibinfo{author}{\bibfnamefont{G.}~\bibnamefont{Mussler}},
  \bibinfo{author}{\bibfnamefont{D.}~\bibnamefont{Gr\"{u}tzmacher}},
  \bibnamefont{and} \bibinfo{author}{\bibfnamefont{J.~C.} \bibnamefont{Greer}},
  \bibinfo{journal}{Appl.~Phys.~Lett.} \textbf{\bibinfo{volume}{110}},
  \bibinfo{pages}{093111} (\bibinfo{year}{2017}).

\bibitem[{\citenamefont{Gity et~al.}(2018)\citenamefont{Gity, Ansari,
  K\"{o}nig, Verni, Holmes, Long, Lanius, Sch\"{u}ffelgen, Mussler,
  Gr\"{u}tzmacher et~al.}}]{GAK18}
\bibinfo{author}{\bibfnamefont{F.}~\bibnamefont{Gity}},
  \bibinfo{author}{\bibfnamefont{L.}~\bibnamefont{Ansari}},
  \bibinfo{author}{\bibfnamefont{C.}~\bibnamefont{K\"{o}nig}},
  \bibinfo{author}{\bibfnamefont{G.~A.} \bibnamefont{Verni}},
  \bibinfo{author}{\bibfnamefont{J.~D.} \bibnamefont{Holmes}},
  \bibinfo{author}{\bibfnamefont{B.}~\bibnamefont{Long}},
  \bibinfo{author}{\bibfnamefont{M.}~\bibnamefont{Lanius}},
  \bibinfo{author}{\bibfnamefont{P.}~\bibnamefont{Sch\"{u}ffelgen}},
  \bibinfo{author}{\bibfnamefont{G.}~\bibnamefont{Mussler}},
  \bibinfo{author}{\bibfnamefont{D.}~\bibnamefont{Gr\"{u}tzmacher}},
  \bibnamefont{et~al.}, \bibinfo{journal}{Microelectron.~Eng.}
  \textbf{\bibinfo{volume}{195}}, \bibinfo{pages}{21} (\bibinfo{year}{2018}).

\bibitem[{\citenamefont{Greer et~al.}(2018)\citenamefont{Greer, Blom, and
  Ansari}}]{GBA18}
\bibinfo{author}{\bibfnamefont{J.~C.} \bibnamefont{Greer}},
  \bibinfo{author}{\bibfnamefont{A.}~\bibnamefont{Blom}}, \bibnamefont{and}
  \bibinfo{author}{\bibfnamefont{L.}~\bibnamefont{Ansari}},
  \bibinfo{journal}{J.~Phys.: Condens.~Matter} \textbf{\bibinfo{volume}{30}},
  \bibinfo{pages}{414003} (\bibinfo{year}{2018}).

\bibitem[{\citenamefont{Cohen and van Lieshout}(1935)}]{Cohen1935}
\bibinfo{author}{\bibfnamefont{E.}~\bibnamefont{Cohen}} \bibnamefont{and}
  \bibinfo{author}{\bibfnamefont{A.~K.~W.~A.} \bibnamefont{van Lieshout}},
  \bibinfo{journal}{Z.~Phys.~Chem.~(N.~F.)} \textbf{\bibinfo{volume}{173A}}
  (\bibinfo{year}{1935}).

\bibitem[{\citenamefont{Houben et~al.}(2019)\citenamefont{Houben, Jochum,
  Lozano, Bisht, Men\'{e}ndez, Merkel, R\"{u}ffer, Chumakov, Roelants, Partoens
  et~al.}}]{Houben_PRB_2019}
\bibinfo{author}{\bibfnamefont{K.}~\bibnamefont{Houben}},
  \bibinfo{author}{\bibfnamefont{J.~K.} \bibnamefont{Jochum}},
  \bibinfo{author}{\bibfnamefont{D.~P.} \bibnamefont{Lozano}},
  \bibinfo{author}{\bibfnamefont{M.}~\bibnamefont{Bisht}},
  \bibinfo{author}{\bibfnamefont{E.}~\bibnamefont{Men\'{e}ndez}},
  \bibinfo{author}{\bibfnamefont{D.~G.} \bibnamefont{Merkel}},
  \bibinfo{author}{\bibfnamefont{R.}~\bibnamefont{R\"{u}ffer}},
  \bibinfo{author}{\bibfnamefont{A.~I.} \bibnamefont{Chumakov}},
  \bibinfo{author}{\bibfnamefont{S.}~\bibnamefont{Roelants}},
  \bibinfo{author}{\bibfnamefont{B.}~\bibnamefont{Partoens}},
  \bibnamefont{et~al.}, \bibinfo{journal}{Phys.~Rev.~B}
  \textbf{\bibinfo{volume}{100}}, \bibinfo{pages}{075408}
  (\bibinfo{year}{2019}).

\bibitem[{\citenamefont{Farrow et~al.}(1981)\citenamefont{Farrow, Robertson,
  Williams, Cullis, Jones, Young, and Dennis}}]{Farrow1981}
\bibinfo{author}{\bibfnamefont{R.~F.~C.} \bibnamefont{Farrow}},
  \bibinfo{author}{\bibfnamefont{D.~S.} \bibnamefont{Robertson}},
  \bibinfo{author}{\bibfnamefont{G.~M.} \bibnamefont{Williams}},
  \bibinfo{author}{\bibfnamefont{A.~G.} \bibnamefont{Cullis}},
  \bibinfo{author}{\bibfnamefont{G.~R.} \bibnamefont{Jones}},
  \bibinfo{author}{\bibfnamefont{I.~M.} \bibnamefont{Young}}, \bibnamefont{and}
  \bibinfo{author}{\bibfnamefont{P.~N.~J.} \bibnamefont{Dennis}},
  \bibinfo{journal}{J.~Cryst.~Growth} \textbf{\bibinfo{volume}{54}},
  \bibinfo{pages}{507} (\bibinfo{year}{1981}).

\bibitem[{\citenamefont{Bowman et~al.}(1990)\citenamefont{Bowman, Adams,
  Engelhart, and H\"{o}chst}}]{Bowman1990}
\bibinfo{author}{\bibfnamefont{R.~C.} \bibnamefont{Bowman}},
  \bibinfo{author}{\bibfnamefont{P.~M.} \bibnamefont{Adams}},
  \bibinfo{author}{\bibfnamefont{M.~A.} \bibnamefont{Engelhart}},
  \bibnamefont{and}
  \bibinfo{author}{\bibfnamefont{H.}~\bibnamefont{H\"{o}chst}},
  \bibinfo{journal}{J.~Vac.~Sci.~Technol.~A} \textbf{\bibinfo{volume}{8}},
  \bibinfo{pages}{1577} (\bibinfo{year}{1990}).

\bibitem[{\citenamefont{Groves and Paul}(1963)}]{Groves1963}
\bibinfo{author}{\bibfnamefont{S.}~\bibnamefont{Groves}} \bibnamefont{and}
  \bibinfo{author}{\bibfnamefont{W.}~\bibnamefont{Paul}},
  \bibinfo{journal}{Phys.~Rev.~Lett.} \textbf{\bibinfo{volume}{11}},
  \bibinfo{pages}{194} (\bibinfo{year}{1963}).

\bibitem[{\citenamefont{K\"{u}fner et~al.}(2013)\citenamefont{K\"{u}fner,
  Furthm\"{u}ller, Matthes, Fitzner, and Bechstedt}}]{Kuefner2013}
\bibinfo{author}{\bibfnamefont{S.}~\bibnamefont{K\"{u}fner}},
  \bibinfo{author}{\bibfnamefont{J.}~\bibnamefont{Furthm\"{u}ller}},
  \bibinfo{author}{\bibfnamefont{L.}~\bibnamefont{Matthes}},
  \bibinfo{author}{\bibfnamefont{M.}~\bibnamefont{Fitzner}}, \bibnamefont{and}
  \bibinfo{author}{\bibfnamefont{F.}~\bibnamefont{Bechstedt}},
  \bibinfo{journal}{Phys.~Rev.~B} \textbf{\bibinfo{volume}{87}},
  \bibinfo{pages}{235307} (\bibinfo{year}{2013}).

\bibitem[{\citenamefont{Lan et~al.}(2017)\citenamefont{Lan, Chang, and
  Liu}}]{Lan2017}
\bibinfo{author}{\bibfnamefont{H.-S.} \bibnamefont{Lan}},
  \bibinfo{author}{\bibfnamefont{S.~T.} \bibnamefont{Chang}}, \bibnamefont{and}
  \bibinfo{author}{\bibfnamefont{C.~W.} \bibnamefont{Liu}},
  \bibinfo{journal}{Phys.~Rev.~B} \textbf{\bibinfo{volume}{95}},
  \bibinfo{pages}{201201} (\bibinfo{year}{2017}).

\bibitem[{\citenamefont{Ewald}(1954)}]{Ewald1954}
\bibinfo{author}{\bibfnamefont{A.~W.} \bibnamefont{Ewald}},
  \bibinfo{journal}{J.~Appl.~Phys.} \textbf{\bibinfo{volume}{25}},
  \bibinfo{pages}{1436} (\bibinfo{year}{1954}).

\bibitem[{\citenamefont{Olesinski and
  Abbaschian}(1984)}]{OlesinskiAbbaschian1984}
\bibinfo{author}{\bibfnamefont{R.~W.} \bibnamefont{Olesinski}}
  \bibnamefont{and} \bibinfo{author}{\bibfnamefont{G.~J.}
  \bibnamefont{Abbaschian}}, \bibinfo{journal}{Bull.~Alloy Phase.~Diagr.}
  \textbf{\bibinfo{volume}{5}}, \bibinfo{pages}{265} (\bibinfo{year}{1984}).

\bibitem[{\citenamefont{Piao et~al.}(1990)\citenamefont{Piao, Beresford,
  Licata, Wang, and Homma}}]{Piao1990}
\bibinfo{author}{\bibfnamefont{J.}~\bibnamefont{Piao}},
  \bibinfo{author}{\bibfnamefont{R.}~\bibnamefont{Beresford}},
  \bibinfo{author}{\bibfnamefont{T.}~\bibnamefont{Licata}},
  \bibinfo{author}{\bibfnamefont{W.~I.} \bibnamefont{Wang}}, \bibnamefont{and}
  \bibinfo{author}{\bibfnamefont{H.}~\bibnamefont{Homma}},
  \bibinfo{journal}{J.~Vac.~Sci.~Technol.~B} \textbf{\bibinfo{volume}{8}},
  \bibinfo{pages}{221} (\bibinfo{year}{1990}).

\bibitem[{\citenamefont{He and Atwater}(1996)}]{HeAtwater1996}
\bibinfo{author}{\bibfnamefont{G.}~\bibnamefont{He}} \bibnamefont{and}
  \bibinfo{author}{\bibfnamefont{H.~A.} \bibnamefont{Atwater}},
  \bibinfo{journal}{Appl.~Phys.~Lett.} \textbf{\bibinfo{volume}{68}},
  \bibinfo{pages}{664} (\bibinfo{year}{1996}).

\bibitem[{\citenamefont{Biswas et~al.}(2016)\citenamefont{Biswas, Doherty,
  Saladukha, Ramasse, Majumdar, Upmanyu, Singha, Ochalski, Morris, and
  Holmes}}]{Biswas2016}
\bibinfo{author}{\bibfnamefont{S.}~\bibnamefont{Biswas}},
  \bibinfo{author}{\bibfnamefont{J.}~\bibnamefont{Doherty}},
  \bibinfo{author}{\bibfnamefont{D.}~\bibnamefont{Saladukha}},
  \bibinfo{author}{\bibfnamefont{Q.}~\bibnamefont{Ramasse}},
  \bibinfo{author}{\bibfnamefont{D.}~\bibnamefont{Majumdar}},
  \bibinfo{author}{\bibfnamefont{M.}~\bibnamefont{Upmanyu}},
  \bibinfo{author}{\bibfnamefont{A.}~\bibnamefont{Singha}},
  \bibinfo{author}{\bibfnamefont{T.}~\bibnamefont{Ochalski}},
  \bibinfo{author}{\bibfnamefont{M.~A.} \bibnamefont{Morris}},
  \bibnamefont{and} \bibinfo{author}{\bibfnamefont{J.~D.}
  \bibnamefont{Holmes}}, \bibinfo{journal}{Nature Communications}
  \textbf{\bibinfo{volume}{7}}, \bibinfo{pages}{11405} (\bibinfo{year}{2016}).

\bibitem[{\citenamefont{Suzuki et~al.}(2016)\citenamefont{Suzuki, Nakatsuka,
  Shibayama, Sakashita, Takeuchi, Kurosawa, and Zaima}}]{Suzuki2016}
\bibinfo{author}{\bibfnamefont{A.}~\bibnamefont{Suzuki}},
  \bibinfo{author}{\bibfnamefont{O.}~\bibnamefont{Nakatsuka}},
  \bibinfo{author}{\bibfnamefont{S.}~\bibnamefont{Shibayama}},
  \bibinfo{author}{\bibfnamefont{M.}~\bibnamefont{Sakashita}},
  \bibinfo{author}{\bibfnamefont{W.}~\bibnamefont{Takeuchi}},
  \bibinfo{author}{\bibfnamefont{M.}~\bibnamefont{Kurosawa}}, \bibnamefont{and}
  \bibinfo{author}{\bibfnamefont{S.}~\bibnamefont{Zaima}},
  \bibinfo{journal}{Jpn.~J.~Appl.~Phys.} \textbf{\bibinfo{volume}{55}},
  \bibinfo{pages}{04EB12} (\bibinfo{year}{2016}).

\bibitem[{\citenamefont{Sanchez-Soares
  et~al.}(2019)\citenamefont{Sanchez-Soares, O'Donnell, and
  Greer}}]{Miscibility}
\bibinfo{author}{\bibfnamefont{A.}~\bibnamefont{Sanchez-Soares}},
  \bibinfo{author}{\bibfnamefont{C.}~\bibnamefont{O'Donnell}},
  \bibnamefont{and} \bibinfo{author}{\bibfnamefont{J.~C.} \bibnamefont{Greer}},
  \bibinfo{journal}{arXiv:1904.09147}  (\bibinfo{year}{2019}),
  \urlprefix\url{http://arxiv.org/abs/1904.09147}.

\bibitem[{\citenamefont{Gupta et~al.}(2013)\citenamefont{Gupta,
  Magyari-K\"{o}pe, Nishi, and Saraswat}}]{Gupta2013}
\bibinfo{author}{\bibfnamefont{S.}~\bibnamefont{Gupta}},
  \bibinfo{author}{\bibfnamefont{B.}~\bibnamefont{Magyari-K\"{o}pe}},
  \bibinfo{author}{\bibfnamefont{Y.}~\bibnamefont{Nishi}}, \bibnamefont{and}
  \bibinfo{author}{\bibfnamefont{K.~C.} \bibnamefont{Saraswat}},
  \bibinfo{journal}{J.~Appl.~Phys.} \textbf{\bibinfo{volume}{113}},
  \bibinfo{pages}{073707} (\bibinfo{year}{2013}).

\bibitem[{\citenamefont{Virgilio et~al.}(2013)\citenamefont{Virgilio,
  Manganelli, Grosso, Pizzi, and Capellini}}]{Virgilio2013}
\bibinfo{author}{\bibfnamefont{M.}~\bibnamefont{Virgilio}},
  \bibinfo{author}{\bibfnamefont{C.~L.} \bibnamefont{Manganelli}},
  \bibinfo{author}{\bibfnamefont{G.}~\bibnamefont{Grosso}},
  \bibinfo{author}{\bibfnamefont{G.}~\bibnamefont{Pizzi}}, \bibnamefont{and}
  \bibinfo{author}{\bibfnamefont{G.}~\bibnamefont{Capellini}},
  \bibinfo{journal}{Phys.~Rev.~B} \textbf{\bibinfo{volume}{87}},
  \bibinfo{pages}{235313} (\bibinfo{year}{2013}).

\bibitem[{\citenamefont{Kurdi et~al.}(2010{\natexlab{a}})\citenamefont{Kurdi,
  Fishman, Sauvage, and Boucaud}}]{ElKurdi2010a}
\bibinfo{author}{\bibfnamefont{M.~E.} \bibnamefont{Kurdi}},
  \bibinfo{author}{\bibfnamefont{G.}~\bibnamefont{Fishman}},
  \bibinfo{author}{\bibfnamefont{S.}~\bibnamefont{Sauvage}}, \bibnamefont{and}
  \bibinfo{author}{\bibfnamefont{P.}~\bibnamefont{Boucaud}},
  \bibinfo{journal}{J.~Appl.~Phys.} \textbf{\bibinfo{volume}{107}},
  \bibinfo{pages}{013710} (\bibinfo{year}{2010}{\natexlab{a}}).

\bibitem[{\citenamefont{O'Halloran et~al.}(2019)\citenamefont{O'Halloran,
  Broderick, Tanner, Schulz, and O'Reilly}}]{OHalloran2019}
\bibinfo{author}{\bibfnamefont{E.~J.} \bibnamefont{O'Halloran}},
  \bibinfo{author}{\bibfnamefont{C.~A.} \bibnamefont{Broderick}},
  \bibinfo{author}{\bibfnamefont{D.~S.~P.} \bibnamefont{Tanner}},
  \bibinfo{author}{\bibfnamefont{S.}~\bibnamefont{Schulz}}, \bibnamefont{and}
  \bibinfo{author}{\bibfnamefont{E.~P.} \bibnamefont{O'Reilly}},
  \bibinfo{journal}{Opt.~Quantum Electron.} \textbf{\bibinfo{volume}{51}},
  \bibinfo{pages}{314} (\bibinfo{year}{2019}).

\bibitem[{\citenamefont{Tanner et~al.}(2020)\citenamefont{Tanner, Broderick,
  Kirwan, Schulz, and O'Reilly}}]{Tanner_submitted_2020}
\bibinfo{author}{\bibfnamefont{D.~S.~P.} \bibnamefont{Tanner}},
  \bibinfo{author}{\bibfnamefont{C.~A.} \bibnamefont{Broderick}},
  \bibinfo{author}{\bibfnamefont{A.~C.} \bibnamefont{Kirwan}},
  \bibinfo{author}{\bibfnamefont{S.}~\bibnamefont{Schulz}}, \bibnamefont{and}
  \bibinfo{author}{\bibfnamefont{E.~P.} \bibnamefont{O'Reilly}},
  \bibinfo{journal}{{``Fully analytic valence force fields for the relaxation
  of group-IV semiconductor alloys: elastic properties of group-IV materials
  calculated from first principles'', submitted}}  (\bibinfo{year}{2020}).

\bibitem[{\citenamefont{Dreizler}(1990)}]{Dreizler1990d}
\bibinfo{author}{\bibfnamefont{E.~K.~U.} \bibnamefont{Dreizler},
  \bibfnamefont{Reiner M.~and~Gross}}, in \emph{\bibinfo{booktitle}{Density
  Functional Theory}} (\bibinfo{publisher}{Springer}, \bibinfo{year}{1990}),
  ISBN \bibinfo{isbn}{978-3-642-86107-9}.

\bibitem[{\citenamefont{Perdew and Zunger}(1981)}]{Perdew1981}
\bibinfo{author}{\bibfnamefont{J.~P.} \bibnamefont{Perdew}} \bibnamefont{and}
  \bibinfo{author}{\bibfnamefont{A.}~\bibnamefont{Zunger}},
  \bibinfo{journal}{Phys.~Rev.~B} \textbf{\bibinfo{volume}{23}},
  \bibinfo{pages}{5048} (\bibinfo{year}{1981}).

\bibitem[{\citenamefont{Tran and Blaha}(2009)}]{Tran2009}
\bibinfo{author}{\bibfnamefont{F.}~\bibnamefont{Tran}} \bibnamefont{and}
  \bibinfo{author}{\bibfnamefont{P.}~\bibnamefont{Blaha}},
  \bibinfo{journal}{Phys.~Rev.~Lett.} \textbf{\bibinfo{volume}{102}},
  \bibinfo{pages}{226401} (\bibinfo{year}{2009}).

\bibitem[{\citenamefont{Soler et~al.}(2002)\citenamefont{Soler, Artacho, Gale,
  Garc\'{i}a, Junquera, Ordej\'{o}n, and S\'{a}nchez-Portal}}]{Soler2002}
\bibinfo{author}{\bibfnamefont{J.~M.} \bibnamefont{Soler}},
  \bibinfo{author}{\bibfnamefont{E.}~\bibnamefont{Artacho}},
  \bibinfo{author}{\bibfnamefont{J.~D.} \bibnamefont{Gale}},
  \bibinfo{author}{\bibfnamefont{A.}~\bibnamefont{Garc\'{i}a}},
  \bibinfo{author}{\bibfnamefont{J.}~\bibnamefont{Junquera}},
  \bibinfo{author}{\bibfnamefont{P.}~\bibnamefont{Ordej\'{o}n}},
  \bibnamefont{and}
  \bibinfo{author}{\bibfnamefont{D.}~\bibnamefont{S\'{a}nchez-Portal}},
  \bibinfo{journal}{J.~Phys.: Condens.~Matter} \textbf{\bibinfo{volume}{14}},
  \bibinfo{pages}{2745} (\bibinfo{year}{2002}).

\bibitem[{\citenamefont{Ozaki}(2003)}]{Ozaki2003}
\bibinfo{author}{\bibfnamefont{T.}~\bibnamefont{Ozaki}},
  \bibinfo{journal}{Phys.~Rev.~B} \textbf{\bibinfo{volume}{67}},
  \bibinfo{pages}{155108} (\bibinfo{year}{2003}).

\bibitem[{\citenamefont{Ozaki and Kino}(2004)}]{Ozaki2004}
\bibinfo{author}{\bibfnamefont{T.}~\bibnamefont{Ozaki}} \bibnamefont{and}
  \bibinfo{author}{\bibfnamefont{H.}~\bibnamefont{Kino}},
  \bibinfo{journal}{Phys.~Rev.~B} \textbf{\bibinfo{volume}{69}},
  \bibinfo{pages}{195113} (\bibinfo{year}{2004}).

\bibitem[{\citenamefont{Monkhorst and Pack}(1976)}]{Monkhorst1976}
\bibinfo{author}{\bibfnamefont{H.~J.} \bibnamefont{Monkhorst}}
  \bibnamefont{and} \bibinfo{author}{\bibfnamefont{J.~D.} \bibnamefont{Pack}},
  \bibinfo{journal}{Phys.~Rev.~B} \textbf{\bibinfo{volume}{13}},
  \bibinfo{pages}{5188} (\bibinfo{year}{1976}).

\bibitem[{\citenamefont{Eckhardt et~al.}(2014)\citenamefont{Eckhardt, Hummer,
  and Kresse}}]{Eckhardt_PRB_2014}
\bibinfo{author}{\bibfnamefont{C.}~\bibnamefont{Eckhardt}},
  \bibinfo{author}{\bibfnamefont{K.}~\bibnamefont{Hummer}}, \bibnamefont{and}
  \bibinfo{author}{\bibfnamefont{G.}~\bibnamefont{Kresse}},
  \bibinfo{journal}{Phys.~Rev.~B} \textbf{\bibinfo{volume}{89}},
  \bibinfo{pages}{165201} (\bibinfo{year}{2014}).

\bibitem[{\citenamefont{Baker and Hart}(1975)}]{Baker1975}
\bibinfo{author}{\bibfnamefont{J.~F.~C.} \bibnamefont{Baker}} \bibnamefont{and}
  \bibinfo{author}{\bibfnamefont{M.}~\bibnamefont{Hart}},
  \bibinfo{journal}{Acta Crystallogr.~A} \textbf{\bibinfo{volume}{31}},
  \bibinfo{pages}{364} (\bibinfo{year}{1975}).

\bibitem[{\citenamefont{Thewlis and Davey}(1954)}]{Thewlis1954}
\bibinfo{author}{\bibfnamefont{J.}~\bibnamefont{Thewlis}} \bibnamefont{and}
  \bibinfo{author}{\bibfnamefont{A.~R.} \bibnamefont{Davey}},
  \bibinfo{journal}{Nature} \textbf{\bibinfo{volume}{174}},
  \bibinfo{pages}{1011} (\bibinfo{year}{1954}).

\bibitem[{\citenamefont{McSkimin and Andreatch}(1963)}]{McSkimin1963}
\bibinfo{author}{\bibfnamefont{H.~J.} \bibnamefont{McSkimin}} \bibnamefont{and}
  \bibinfo{author}{\bibfnamefont{P.}~\bibnamefont{Andreatch}},
  \bibinfo{journal}{J.~Appl.~Phys.} \textbf{\bibinfo{volume}{34}},
  \bibinfo{pages}{651} (\bibinfo{year}{1963}).

\bibitem[{\citenamefont{Price et~al.}(1971)\citenamefont{Price, Rowe, and
  Nicklow}}]{Price1971}
\bibinfo{author}{\bibfnamefont{D.~L.} \bibnamefont{Price}},
  \bibinfo{author}{\bibfnamefont{J.~M.} \bibnamefont{Rowe}}, \bibnamefont{and}
  \bibinfo{author}{\bibfnamefont{R.~M.} \bibnamefont{Nicklow}},
  \bibinfo{journal}{Phys.~Rev.~B} \textbf{\bibinfo{volume}{3}},
  \bibinfo{pages}{1268} (\bibinfo{year}{1971}).

\bibitem[{\citenamefont{Shen}(1994)}]{Shen1994}
\bibinfo{author}{\bibfnamefont{S.-G.} \bibnamefont{Shen}},
  \bibinfo{journal}{J.~Phys.: Condens.~Matter} \textbf{\bibinfo{volume}{6}},
  \bibinfo{pages}{8733} (\bibinfo{year}{1994}).

\bibitem[{\citenamefont{Huntington}(1958)}]{Huntington1958}
\bibinfo{author}{\bibfnamefont{H.~B.} \bibnamefont{Huntington}}, in
  \emph{\bibinfo{booktitle}{Solid State Physics}}
  (\bibinfo{publisher}{Elsevier}, \bibinfo{year}{1958}),
  vol.~\bibinfo{volume}{7}, pp. \bibinfo{pages}{213--351}.

\bibitem[{\citenamefont{Zdetsis}(1977)}]{Zdetsis1977}
\bibinfo{author}{\bibfnamefont{A.~D.} \bibnamefont{Zdetsis}},
  \bibinfo{journal}{J.~Phys.~Chem.~Solids} \textbf{\bibinfo{volume}{38}},
  \bibinfo{pages}{1113} (\bibinfo{year}{1977}).

\bibitem[{\citenamefont{Zwerdling et~al.}(1959)\citenamefont{Zwerdling, Lax,
  Roth, and Button}}]{Zwerdling1959}
\bibinfo{author}{\bibfnamefont{S.}~\bibnamefont{Zwerdling}},
  \bibinfo{author}{\bibfnamefont{B.}~\bibnamefont{Lax}},
  \bibinfo{author}{\bibfnamefont{L.~M.} \bibnamefont{Roth}}, \bibnamefont{and}
  \bibinfo{author}{\bibfnamefont{K.~J.} \bibnamefont{Button}},
  \bibinfo{journal}{Phys.~Rev.} \textbf{\bibinfo{volume}{114}},
  \bibinfo{pages}{80} (\bibinfo{year}{1959}).

\bibitem[{\citenamefont{Booth and Ewald}(1968)}]{Booth1968_a}
\bibinfo{author}{\bibfnamefont{B.~L.} \bibnamefont{Booth}} \bibnamefont{and}
  \bibinfo{author}{\bibfnamefont{A.~W.} \bibnamefont{Ewald}},
  \bibinfo{journal}{Phys.~Rev.} \textbf{\bibinfo{volume}{168}},
  \bibinfo{pages}{796} (\bibinfo{year}{1968}).

\bibitem[{\citenamefont{Wirths et~al.}(2016)\citenamefont{Wirths, Buca, and
  Mantl}}]{WirthsBucaMantl2016}
\bibinfo{author}{\bibfnamefont{S.}~\bibnamefont{Wirths}},
  \bibinfo{author}{\bibfnamefont{D.}~\bibnamefont{Buca}}, \bibnamefont{and}
  \bibinfo{author}{\bibfnamefont{S.}~\bibnamefont{Mantl}},
  \bibinfo{journal}{Prog.~Cryst.~Growth Charact.~Mater.}
  \textbf{\bibinfo{volume}{62}}, \bibinfo{pages}{1} (\bibinfo{year}{2016}).

\bibitem[{\citenamefont{Groves et~al.}(1970)\citenamefont{Groves, Pidgeon,
  Ewald, and Wagner}}]{GROVES19702031}
\bibinfo{author}{\bibfnamefont{S.~H.} \bibnamefont{Groves}},
  \bibinfo{author}{\bibfnamefont{C.~R.} \bibnamefont{Pidgeon}},
  \bibinfo{author}{\bibfnamefont{A.~W.} \bibnamefont{Ewald}}, \bibnamefont{and}
  \bibinfo{author}{\bibfnamefont{R.~J.} \bibnamefont{Wagner}},
  \bibinfo{journal}{J.~Phys.~Chem.~Solids} \textbf{\bibinfo{volume}{31}},
  \bibinfo{pages}{2031} (\bibinfo{year}{1970}).

\bibitem[{\citenamefont{Qteish and Needs}(1992)}]{Qteish1992}
\bibinfo{author}{\bibfnamefont{A.}~\bibnamefont{Qteish}} \bibnamefont{and}
  \bibinfo{author}{\bibfnamefont{R.~J.} \bibnamefont{Needs}},
  \bibinfo{journal}{Phys.~Rev.~B} \textbf{\bibinfo{volume}{45}},
  \bibinfo{pages}{1317} (\bibinfo{year}{1992}).

\bibitem[{\citenamefont{R\"{o}dl et~al.}(2019)\citenamefont{R\"{o}dl,
  Furthm\"{u}ller, Suckert, Armuzza, Bechstedt, and Botti}}]{Roedl}
\bibinfo{author}{\bibfnamefont{C.}~\bibnamefont{R\"{o}dl}},
  \bibinfo{author}{\bibfnamefont{J.}~\bibnamefont{Furthm\"{u}ller}},
  \bibinfo{author}{\bibfnamefont{J.~R.} \bibnamefont{Suckert}},
  \bibinfo{author}{\bibfnamefont{V.}~\bibnamefont{Armuzza}},
  \bibinfo{author}{\bibfnamefont{F.}~\bibnamefont{Bechstedt}},
  \bibnamefont{and} \bibinfo{author}{\bibfnamefont{S.}~\bibnamefont{Botti}},
  \bibinfo{journal}{Phys.~Rev.~Materials} \textbf{\bibinfo{volume}{3}},
  \bibinfo{pages}{034602} (\bibinfo{year}{2019}).

\bibitem[{\citenamefont{Madelung and R\"{o}ssler}(2002)}]{Authors2002}
\bibinfo{author}{\bibfnamefont{O.}~\bibnamefont{Madelung}} \bibnamefont{and}
  \bibinfo{author}{\bibfnamefont{U.}~\bibnamefont{R\"{o}ssler}},
  \emph{\bibinfo{title}{{Group-IV Elements, IV-IV and III-V Compounds.~Part b -
  Electronic, Transport, Optical and Other Properties}}}
  (\bibinfo{publisher}{Springer}, \bibinfo{address}{Heidelberg},
  \bibinfo{year}{2002}).

\bibitem[{\citenamefont{Wei and Zunger}(1999)}]{Wei1999}
\bibinfo{author}{\bibfnamefont{S.-H.} \bibnamefont{Wei}} \bibnamefont{and}
  \bibinfo{author}{\bibfnamefont{A.}~\bibnamefont{Zunger}},
  \bibinfo{journal}{Phys.~Rev.~B} \textbf{\bibinfo{volume}{60}},
  \bibinfo{pages}{5404} (\bibinfo{year}{1999}).

\bibitem[{\citenamefont{de~Walle}(1989)}]{VandeWalle1989}
\bibinfo{author}{\bibfnamefont{C.~G.~V.} \bibnamefont{de~Walle}},
  \bibinfo{journal}{Phys.~Rev.~B} \textbf{\bibinfo{volume}{39}},
  \bibinfo{pages}{1871} (\bibinfo{year}{1989}).

\bibitem[{\citenamefont{Fischetti and Laux}(1996)}]{Fischetti1996}
\bibinfo{author}{\bibfnamefont{M.~V.} \bibnamefont{Fischetti}}
  \bibnamefont{and} \bibinfo{author}{\bibfnamefont{S.~E.} \bibnamefont{Laux}},
  \bibinfo{journal}{J.~Appl.~Phys.} \textbf{\bibinfo{volume}{80}},
  \bibinfo{pages}{2234} (\bibinfo{year}{1996}).

\bibitem[{\citenamefont{Roman and Ewald}(1972)}]{Roman1972}
\bibinfo{author}{\bibfnamefont{B.~J.} \bibnamefont{Roman}} \bibnamefont{and}
  \bibinfo{author}{\bibfnamefont{A.~W.} \bibnamefont{Ewald}},
  \bibinfo{journal}{Phys.~Rev.~B} \textbf{\bibinfo{volume}{5}},
  \bibinfo{pages}{3914} (\bibinfo{year}{1972}).

\bibitem[{\citenamefont{Zunger et~al.}(1990)\citenamefont{Zunger, Wei,
  Ferreira, and Bernard}}]{Zunger1990}
\bibinfo{author}{\bibfnamefont{A.}~\bibnamefont{Zunger}},
  \bibinfo{author}{\bibfnamefont{S.~H.} \bibnamefont{Wei}},
  \bibinfo{author}{\bibfnamefont{L.~G.} \bibnamefont{Ferreira}},
  \bibnamefont{and} \bibinfo{author}{\bibfnamefont{J.~E.}
  \bibnamefont{Bernard}}, \bibinfo{journal}{Phys.~Rev.~Lett.}
  \textbf{\bibinfo{volume}{65}}, \bibinfo{pages}{353} (\bibinfo{year}{1990}).

\bibitem[{\citenamefont{Hass et~al.}(1990)\citenamefont{Hass, Davis, and
  Zunger}}]{Hass1990}
\bibinfo{author}{\bibfnamefont{K.~C.} \bibnamefont{Hass}},
  \bibinfo{author}{\bibfnamefont{L.~C.} \bibnamefont{Davis}}, \bibnamefont{and}
  \bibinfo{author}{\bibfnamefont{A.}~\bibnamefont{Zunger}},
  \bibinfo{journal}{Phys.~Rev.~B} \textbf{\bibinfo{volume}{42}},
  \bibinfo{pages}{3757} (\bibinfo{year}{1990}).

\bibitem[{\citenamefont{Walle et~al.}(2002)\citenamefont{Walle, Asta, and
  Ceder}}]{ATAT2002}
\bibinfo{author}{\bibfnamefont{A.~V.~D.} \bibnamefont{Walle}},
  \bibinfo{author}{\bibfnamefont{.}~\bibnamefont{Asta}}, \bibnamefont{and}
  \bibinfo{author}{\bibfnamefont{G.}~\bibnamefont{Ceder}},
  \bibinfo{journal}{Calphad} \textbf{\bibinfo{volume}{26}},
  \bibinfo{pages}{539} (\bibinfo{year}{2002}).

\bibitem[{\citenamefont{Krijn}(1991)}]{Krijn_SST_1991}
\bibinfo{author}{\bibfnamefont{M.~P.~C.~M.} \bibnamefont{Krijn}},
  \bibinfo{journal}{Semicond.~Sci.~Technol.} \textbf{\bibinfo{volume}{6}},
  \bibinfo{pages}{27} (\bibinfo{year}{1991}).

\bibitem[{\citenamefont{Usman et~al.}(2018)\citenamefont{Usman, Broderick, and
  O'Reilly}}]{Usman_PRA_2018}
\bibinfo{author}{\bibfnamefont{M.}~\bibnamefont{Usman}},
  \bibinfo{author}{\bibfnamefont{C.~A.} \bibnamefont{Broderick}},
  \bibnamefont{and} \bibinfo{author}{\bibfnamefont{E.~P.}
  \bibnamefont{O'Reilly}}, \bibinfo{journal}{Phys.~Rev.~Applied}
  \textbf{\bibinfo{volume}{10}}, \bibinfo{pages}{044024}
  (\bibinfo{year}{2018}).

\bibitem[{\citenamefont{Usman et~al.}(2013)\citenamefont{Usman, Broderick,
  Batool, Hild, Hosea, Sweeney, and O'Reilly}}]{Usman_PRB_2013}
\bibinfo{author}{\bibfnamefont{M.}~\bibnamefont{Usman}},
  \bibinfo{author}{\bibfnamefont{C.~A.} \bibnamefont{Broderick}},
  \bibinfo{author}{\bibfnamefont{Z.}~\bibnamefont{Batool}},
  \bibinfo{author}{\bibfnamefont{K.}~\bibnamefont{Hild}},
  \bibinfo{author}{\bibfnamefont{T.~J.~C.} \bibnamefont{Hosea}},
  \bibinfo{author}{\bibfnamefont{S.~J.} \bibnamefont{Sweeney}},
  \bibnamefont{and} \bibinfo{author}{\bibfnamefont{E.~P.}
  \bibnamefont{O'Reilly}}, \bibinfo{journal}{Phys.~Rev.~B}
  \textbf{\bibinfo{volume}{87}}, \bibinfo{pages}{115104}
  (\bibinfo{year}{2013}).

\bibitem[{\citenamefont{Schulz et~al.}(2018)\citenamefont{Schulz, Broderick,
  O'Halloran, and O'Reilly}}]{Schulz2018}
\bibinfo{author}{\bibfnamefont{S.}~\bibnamefont{Schulz}},
  \bibinfo{author}{\bibfnamefont{C.~A.} \bibnamefont{Broderick}},
  \bibinfo{author}{\bibfnamefont{E.~J.} \bibnamefont{O'Halloran}},
  \bibnamefont{and} \bibinfo{author}{\bibfnamefont{E.~P.}
  \bibnamefont{O'Reilly}}, in \emph{\bibinfo{booktitle}{International
  Conference on Numerical Simulation of Optoelectronic Devices (NUSOD)}}
  (\bibinfo{year}{2018}), pp. \bibinfo{pages}{39--40}.

\bibitem[{\citenamefont{Eales et~al.}(2019)\citenamefont{Eales, Marko, Schulz,
  O'Halloran, Ghetmiri, Du, Zhou, Yu, Margetis, Tolle et~al.}}]{Eales2019}
\bibinfo{author}{\bibfnamefont{T.~D.} \bibnamefont{Eales}},
  \bibinfo{author}{\bibfnamefont{I.~P.} \bibnamefont{Marko}},
  \bibinfo{author}{\bibfnamefont{S.}~\bibnamefont{Schulz}},
  \bibinfo{author}{\bibfnamefont{E.}~\bibnamefont{O'Halloran}},
  \bibinfo{author}{\bibfnamefont{S.}~\bibnamefont{Ghetmiri}},
  \bibinfo{author}{\bibfnamefont{W.}~\bibnamefont{Du}},
  \bibinfo{author}{\bibfnamefont{Y.}~\bibnamefont{Zhou}},
  \bibinfo{author}{\bibfnamefont{S.-Q.} \bibnamefont{Yu}},
  \bibinfo{author}{\bibfnamefont{J.}~\bibnamefont{Margetis}},
  \bibinfo{author}{\bibfnamefont{J.}~\bibnamefont{Tolle}},
  \bibnamefont{et~al.}, \bibinfo{journal}{Sci.~Rep.}
  \textbf{\bibinfo{volume}{9}}, \bibinfo{pages}{1} (\bibinfo{year}{2019}).

\bibitem[{\citenamefont{Caro et~al.}(2012)\citenamefont{Caro, Schulz, and
  O'Reilly}}]{Caro_JPCM_2012}
\bibinfo{author}{\bibfnamefont{M.~A.} \bibnamefont{Caro}},
  \bibinfo{author}{\bibfnamefont{S.}~\bibnamefont{Schulz}}, \bibnamefont{and}
  \bibinfo{author}{\bibfnamefont{E.~P.} \bibnamefont{O'Reilly}},
  \bibinfo{journal}{J.~Phys.: Condens.~Matter} \textbf{\bibinfo{volume}{25}},
  \bibinfo{pages}{025803} (\bibinfo{year}{2012}).

\bibitem[{\citenamefont{O'Reilly}(1989)}]{OReilly1989}
\bibinfo{author}{\bibfnamefont{E.~P.} \bibnamefont{O'Reilly}},
  \bibinfo{journal}{Semicond.~Sci.~Technol.} \textbf{\bibinfo{volume}{4}},
  \bibinfo{pages}{121} (\bibinfo{year}{1989}).

\bibitem[{\citenamefont{Voisin}(1988)}]{Voisin1988}
\bibinfo{author}{\bibfnamefont{P.}~\bibnamefont{Voisin}}, in
  \emph{\bibinfo{booktitle}{{Quantum Wells and Superlattices in Optoelectronic
  Devices and Integrated Optics}}} (\bibinfo{year}{1988}), vol.
  \bibinfo{volume}{861}, pp. \bibinfo{pages}{88--96}.

\bibitem[{\citenamefont{Tomi\'{c} and O'Reilly}(2003)}]{Tomic2003}
\bibinfo{author}{\bibfnamefont{S.}~\bibnamefont{Tomi\'{c}}} \bibnamefont{and}
  \bibinfo{author}{\bibfnamefont{E.~P.} \bibnamefont{O'Reilly}},
  \bibinfo{journal}{IEEE Photo.~Tech.~Lett.} \textbf{\bibinfo{volume}{15}},
  \bibinfo{pages}{6} (\bibinfo{year}{2003}).

\bibitem[{\citenamefont{Broderick et~al.}(2017)\citenamefont{Broderick, Jin,
  Marko, Hild, Ludewig, Bushell, Stolz, Rorison, O'Reilly, Volz
  et~al.}}]{Broderick2017}
\bibinfo{author}{\bibfnamefont{C.~A.} \bibnamefont{Broderick}},
  \bibinfo{author}{\bibfnamefont{S.~R.} \bibnamefont{Jin}},
  \bibinfo{author}{\bibfnamefont{I.~P.} \bibnamefont{Marko}},
  \bibinfo{author}{\bibfnamefont{K.}~\bibnamefont{Hild}},
  \bibinfo{author}{\bibfnamefont{P.}~\bibnamefont{Ludewig}},
  \bibinfo{author}{\bibfnamefont{Z.~L.} \bibnamefont{Bushell}},
  \bibinfo{author}{\bibfnamefont{W.}~\bibnamefont{Stolz}},
  \bibinfo{author}{\bibfnamefont{J.~M.} \bibnamefont{Rorison}},
  \bibinfo{author}{\bibfnamefont{E.~P.} \bibnamefont{O'Reilly}},
  \bibinfo{author}{\bibfnamefont{K.}~\bibnamefont{Volz}}, \bibnamefont{et~al.},
  \bibinfo{journal}{Sci.~Rep.} \textbf{\bibinfo{volume}{7}},
  \bibinfo{pages}{46371} (\bibinfo{year}{2017}).

\bibitem[{\citenamefont{Sanchez-Soares
  et~al.}(2016)\citenamefont{Sanchez-Soares, O'Donnell, and
  Greer}}]{Sanchez-Soares2016a}
\bibinfo{author}{\bibfnamefont{A.}~\bibnamefont{Sanchez-Soares}},
  \bibinfo{author}{\bibfnamefont{C.}~\bibnamefont{O'Donnell}},
  \bibnamefont{and} \bibinfo{author}{\bibfnamefont{J.~C.} \bibnamefont{Greer}},
  \bibinfo{journal}{Phys.~Rev.~B} \textbf{\bibinfo{volume}{94}},
  \bibinfo{pages}{235442} (\bibinfo{year}{2016}).

\bibitem[{\citenamefont{Sanchez et~al.}(1984)\citenamefont{Sanchez, Ducastelle,
  and Gratias}}]{Sanchez1984}
\bibinfo{author}{\bibfnamefont{J.~M.} \bibnamefont{Sanchez}},
  \bibinfo{author}{\bibfnamefont{F.}~\bibnamefont{Ducastelle}},
  \bibnamefont{and} \bibinfo{author}{\bibfnamefont{D.}~\bibnamefont{Gratias}},
  \bibinfo{journal}{Physica A} \textbf{\bibinfo{volume}{128}},
  \bibinfo{pages}{334} (\bibinfo{year}{1984}).

\bibitem[{\citenamefont{Lin et~al.}(2012)\citenamefont{Lin, Chen, Lu, Huo,
  Kamins, and Harris}}]{Lin2012}
\bibinfo{author}{\bibfnamefont{H.}~\bibnamefont{Lin}},
  \bibinfo{author}{\bibfnamefont{R.}~\bibnamefont{Chen}},
  \bibinfo{author}{\bibfnamefont{W.}~\bibnamefont{Lu}},
  \bibinfo{author}{\bibfnamefont{Y.}~\bibnamefont{Huo}},
  \bibinfo{author}{\bibfnamefont{T.}~\bibnamefont{Kamins}}, \bibnamefont{and}
  \bibinfo{author}{\bibfnamefont{J.~S.} \bibnamefont{Harris}},
  \bibinfo{journal}{Appl.~Phys.~Lett.} \textbf{\bibinfo{volume}{100}},
  \bibinfo{pages}{102109} (\bibinfo{year}{2012}).

\bibitem[{\citenamefont{Tran et~al.}(2016)\citenamefont{Tran, Du, Ghetmiri,
  Mosleh, Sun, Soref, Margetis, Tolle, Li, Naseem et~al.}}]{Tran_JAP_2016}
\bibinfo{author}{\bibfnamefont{H.}~\bibnamefont{Tran}},
  \bibinfo{author}{\bibfnamefont{W.}~\bibnamefont{Du}},
  \bibinfo{author}{\bibfnamefont{S.~A.} \bibnamefont{Ghetmiri}},
  \bibinfo{author}{\bibfnamefont{A.}~\bibnamefont{Mosleh}},
  \bibinfo{author}{\bibfnamefont{G.}~\bibnamefont{Sun}},
  \bibinfo{author}{\bibfnamefont{R.~A.} \bibnamefont{Soref}},
  \bibinfo{author}{\bibfnamefont{J.}~\bibnamefont{Margetis}},
  \bibinfo{author}{\bibfnamefont{J.}~\bibnamefont{Tolle}},
  \bibinfo{author}{\bibfnamefont{B.}~\bibnamefont{Li}},
  \bibinfo{author}{\bibfnamefont{H.~A.} \bibnamefont{Naseem}},
  \bibnamefont{et~al.}, \bibinfo{journal}{J.~Appl.~Phys.}
  \textbf{\bibinfo{volume}{119}}, \bibinfo{pages}{103106}
  (\bibinfo{year}{2016}).

\bibitem[{\citenamefont{Xu et~al.}(2019)\citenamefont{Xu, Wallace, Ringwala,
  Chang, Poweleit, Kouvetakis, and Men\'{e}ndez}}]{MenendezA}
\bibinfo{author}{\bibfnamefont{C.}~\bibnamefont{Xu}},
  \bibinfo{author}{\bibfnamefont{P.~M.} \bibnamefont{Wallace}},
  \bibinfo{author}{\bibfnamefont{D.~A.} \bibnamefont{Ringwala}},
  \bibinfo{author}{\bibfnamefont{S.~L.~Y.} \bibnamefont{Chang}},
  \bibinfo{author}{\bibfnamefont{C.~D.} \bibnamefont{Poweleit}},
  \bibinfo{author}{\bibfnamefont{J.}~\bibnamefont{Kouvetakis}},
  \bibnamefont{and}
  \bibinfo{author}{\bibfnamefont{J.}~\bibnamefont{Men\'{e}ndez}},
  \bibinfo{journal}{Appl.~Phys.~Lett.} \textbf{\bibinfo{volume}{114}},
  \bibinfo{pages}{212104} (\bibinfo{year}{2019}).

\bibitem[{\citenamefont{Kurdi et~al.}(2010{\natexlab{b}})\citenamefont{Kurdi,
  Bertin, Martincic, Kersauson, Fishman, Sauvage, Bosseboeuf, and
  Boucaud}}]{ElKurdi2010}
\bibinfo{author}{\bibfnamefont{M.~E.} \bibnamefont{Kurdi}},
  \bibinfo{author}{\bibfnamefont{H.}~\bibnamefont{Bertin}},
  \bibinfo{author}{\bibfnamefont{E.}~\bibnamefont{Martincic}},
  \bibinfo{author}{\bibfnamefont{M.~D.} \bibnamefont{Kersauson}},
  \bibinfo{author}{\bibfnamefont{G.}~\bibnamefont{Fishman}},
  \bibinfo{author}{\bibfnamefont{S.}~\bibnamefont{Sauvage}},
  \bibinfo{author}{\bibfnamefont{A.}~\bibnamefont{Bosseboeuf}},
  \bibnamefont{and} \bibinfo{author}{\bibfnamefont{P.}~\bibnamefont{Boucaud}},
  \bibinfo{journal}{Appl.~Phys.~Lett.} \textbf{\bibinfo{volume}{96}},
  \bibinfo{pages}{041909} (\bibinfo{year}{2010}{\natexlab{b}}).

\bibitem[{\citenamefont{Broderick et~al.}(2019)\citenamefont{Broderick,
  O'Halloran, and O'Reilly}}]{Broderick_arXiv_2019}
\bibinfo{author}{\bibfnamefont{C.~A.} \bibnamefont{Broderick}},
  \bibinfo{author}{\bibfnamefont{E.~J.} \bibnamefont{O'Halloran}},
  \bibnamefont{and} \bibinfo{author}{\bibfnamefont{E.~P.}
  \bibnamefont{O'Reilly}}, \bibinfo{journal}{arXiv:1911.05679}
  (\bibinfo{year}{2019}), \urlprefix\url{http://arxiv.org/abs/1911.05679}.

\bibitem[{\citenamefont{Tsukamoto et~al.}(2015)\citenamefont{Tsukamoto, Hirose,
  Kasamatsu, Mimura, Matsui, and Suda}}]{Tsukamoto2015}
\bibinfo{author}{\bibfnamefont{T.}~\bibnamefont{Tsukamoto}},
  \bibinfo{author}{\bibfnamefont{N.}~\bibnamefont{Hirose}},
  \bibinfo{author}{\bibfnamefont{A.}~\bibnamefont{Kasamatsu}},
  \bibinfo{author}{\bibfnamefont{T.}~\bibnamefont{Mimura}},
  \bibinfo{author}{\bibfnamefont{T.}~\bibnamefont{Matsui}}, \bibnamefont{and}
  \bibinfo{author}{\bibfnamefont{Y.}~\bibnamefont{Suda}},
  \bibinfo{journal}{Appl.~Phys.~Lett.} \textbf{\bibinfo{volume}{106}},
  \bibinfo{pages}{052103} (\bibinfo{year}{2015}).

\bibitem[{\citenamefont{Rainko et~al.}(2020)\citenamefont{Rainko, Ikonic,
  Elbaz, von~den Driesch, Stange, Herth, Boucaud, Kurdi, Gr\"{u}tzmacher, and
  Buca}}]{Rainko_NP_2020}
\bibinfo{author}{\bibfnamefont{D.}~\bibnamefont{Rainko}},
  \bibinfo{author}{\bibfnamefont{Z.}~\bibnamefont{Ikonic}},
  \bibinfo{author}{\bibfnamefont{A.}~\bibnamefont{Elbaz}},
  \bibinfo{author}{\bibfnamefont{N.}~\bibnamefont{von~den Driesch}},
  \bibinfo{author}{\bibfnamefont{D.}~\bibnamefont{Stange}},
  \bibinfo{author}{\bibfnamefont{E.}~\bibnamefont{Herth}},
  \bibinfo{author}{\bibfnamefont{P.}~\bibnamefont{Boucaud}},
  \bibinfo{author}{\bibfnamefont{M.~E.} \bibnamefont{Kurdi}},
  \bibinfo{author}{\bibfnamefont{D.}~\bibnamefont{Gr\"{u}tzmacher}},
  \bibnamefont{and} \bibinfo{author}{\bibfnamefont{D.}~\bibnamefont{Buca}},
  \bibinfo{journal}{Nature Photonics} \textbf{\bibinfo{volume}{14}},
  \bibinfo{pages}{375} (\bibinfo{year}{2020}).

\bibitem[{\citenamefont{Vurgaftman et~al.}(2001)\citenamefont{Vurgaftman,
  Meyer, and Ram-Mohan}}]{Vurgaftman2001}
\bibinfo{author}{\bibfnamefont{I.}~\bibnamefont{Vurgaftman}},
  \bibinfo{author}{\bibfnamefont{J.~R.} \bibnamefont{Meyer}}, \bibnamefont{and}
  \bibinfo{author}{\bibfnamefont{L.~R.} \bibnamefont{Ram-Mohan}},
  \bibinfo{journal}{J.~Appl.~Phys.} \textbf{\bibinfo{volume}{89}},
  \bibinfo{pages}{5815} (\bibinfo{year}{2001}).

\bibitem[{\citenamefont{Haynes}(2014)}]{Haynes2014}
\bibinfo{author}{\bibfnamefont{W.~M.} \bibnamefont{Haynes}},
  \emph{\bibinfo{title}{{CRC Handbook of Chemistry and Physics, 95th Edition}}}
  (\bibinfo{publisher}{CRC Press}, \bibinfo{year}{2014}), ISBN
  \bibinfo{isbn}{978-1482208672}.

\end{thebibliography}

\end{document}